\documentclass[a4paper,11pt]{article}
\pdfoutput=1

\usepackage{jheppub}
\usepackage{booktabs}
\usepackage{graphicx}
\usepackage{amsmath}
\usepackage{amssymb}
\usepackage{amsfonts}
\usepackage{dcolumn}
\usepackage{bm}
\usepackage[dvipsnames]{xcolor}
\usepackage[utf8]{inputenc}
\usepackage{subfigure}
\usepackage{gensymb}
\usepackage{cleveref}
\usepackage{colortbl}
\usepackage{tabularx}
\usepackage{tabulary}
\usepackage{tabularray}
\usepackage{tablefootnote}

\usepackage[T1]{fontenc}
\usepackage{pstricks}
\usepackage{color}
\usepackage{multirow}
\usepackage{slashed}
\usepackage{mathtools}
\usepackage{mathrsfs}
\usepackage{bbold}
\usepackage[force]{feynmp-auto}
\allowdisplaybreaks[4]

\usepackage{verbatim}

\DeclareMathAlphabet{\mathpzc}{OT1}{pzc}{m}{it}

\hypersetup{colorlinks,linkcolor={blue},citecolor={teal},urlcolor={violet}}

\newcommand{\SOTON}{Department of Physics and Astronomy, University of Southampton, SO17 1BJ Southampton, United Kingdom}
\newcommand{\calA}{\mathcal{A}}
\newcommand{\calAS}{\mathcal{A}(S,\overline{S})}

\begin{document}

\title{Modulus stabilisation in the multiple-modulus framework}

\author[1]{Stephen F. King\note{\url{https://orcid.org/0000-0002-4351-7507}}}
\author{and}

\author[2]{Xin Wang\note{\url{https://orcid.org/0000-0003-4292-460X}}}

\affiliation[]{\SOTON}

\emailAdd{king@soton.ac.uk}
\emailAdd{Xin.Wang@soton.ac.uk}

\date{\today}

\abstract{
In a class of modular-invariant models with multiple moduli fields, the viable lepton flavour mixing pattern can be realised if the values of moduli are selected to be at the fixed points. In this paper, we investigate a modulus stabilisation mechanism in the multiple-modulus framework which is capable of providing de Sitter (dS) minima precisely at the fixed points $\tau = {\rm i}$ and $\omega$, by taking into consideration non-perturbative effects on the superpotential and the dilaton K\"ahler potential. Due to the existence of additional K\"ahler moduli, more possible vacua can occur, and the dS vacua could be the deepest under certain conditions. We classify different choices of vacua, and discuss their phenomenological implications for lepton masses and flavour mixing.

}

\maketitle

\section{Introduction}

The flavour problem, that of the origin of the three quark and lepton families and their pattern of masses and mixings, is an unresolved puzzle within the Standard Model (SM) of particle physics. The discovery of very small neutrino masses with large mixing, enriches the flavour problem still further, requiring a further seven parameters (more or less) for its phenomenological description and demanding new physics beyond the SM. The unexpected phenomenon of large lepton mixing has caused a schism in the community between those who think that this is a hint of a family symmetry at work, in particular non-Abelian and discrete, and those who think that it is just a random or anarchic choice of parameters. If one follows the symmetry approach, one is immediately confronted by the problem of how to break the symmetry, without which there would be massless fermions with no mixing, and this leads to the introduction of rather arbitrary flavon fields and driving fields which determine their vacuum alignments which play a crucial role in determining the masses and mixings (for a review see, e.g., Ref.~\cite{King:2013eh}).

In an attempt to make the non-Abelian discrete family symmetries, and in particular the accompanying flavon fields, less arbitrary,
it has been suggested that a more satisfactory framework for addressing the flavour problem, at least in the lepton sector, might be modular symmetry broken by a single complex modulus field $\tau$~\cite{Feruglio:2017spp}. 
Using ideas borrowed from string theory~\cite{Ferrara:1989bc,Ferrara:1989qb},
modular symmetry on the worldsheet represents a reparameterisation symmetry of the extra dimensional coordinates, whose toroidal compactification is controlled by one or more moduli fields, the simplest example being a single complex modulus field $\tau$ describing the two compact dimensional lattice of a six-dimensional theory,
modulus field $\tau$, where its vacuum expectation value (VEV) fixes the geometry of the torus~\cite{Ishiguro:2021ccl, Cremades:2004wa, Ishiguro:2020tmo}. 

The resulting infinite modular symmetry in the upper half of the complex plane, ${\rm PSL}(2,\mathbb{Z})$, has particularly nice features which rely on holomorphicity, the lack of complex conjugation symmetry, which seems to call for supersymmetry.
The infinite modular group has a series of infinite normal subgroups called the principle congruence subgroups $\Gamma (N)$ of level $N$, whose elements are equal to the $2\times 2 $ unit matrix mod $N$ (where typically $N$ is an integer called the level of the group). 
For a given choice of level $N>2$, the quotient group $\Gamma_N={\rm PSL}(2,\mathbb{Z})/\Gamma (N)$ is finite and may be identified with the groups $\Gamma_N=A_4$~\cite{Feruglio:2017spp,Kobayashi:2018vbk, Criado:2018thu, Kobayashi:2018scp,deAnda:2018ecu,Okada:2018yrn,Okada:2019uoy,Nomura:2019yft, Ding:2019gof, Ding:2019zxk,Zhang:2019ngf,Kobayashi:2019gtp,Wang:2019xbo,Okada:2020rjb,Yao:2020qyy,Chen:2021zty, Kobayashi:2021pav, Kang:2022psa, CentellesChulia:2023zhu, Kumar:2023moh}, $S_4$~\cite{Penedo:2018nmg,Novichkov:2018ovf,Kobayashi:2019mna,Wang:2019ovr}, $A_5$~\cite{Novichkov:2018nkm,Ding:2019xna,Criado:2019tzk} for levels $N=3,4,5$, which may subsequently be used as a family symmetry~\cite{Feruglio:2017spp}. 

The only flavon present in such theories is the single modulus field $\tau$, whose VEV fixes the value of Yukawa couplings which form representations of $\Gamma_N$ and
are modular forms. Remarkably, the resulting Yukawa couplings involved in the terms in the superpotential containing superfields whose modular weights do not sum to zero, but take even values,
can exist as modular forms with a precise functional dependence on 
$\tau$~\cite{Feruglio:2017spp}, leading to very predictive theories independent of flavons~\cite{Feruglio:2017spp}.
However, for general values of the modulus field $\tau$, 
the resulting Yukawa couplings are not very hierarchical, so fermion mass hierarchies do not emerge naturally.
There are also more general formulations involving the double cover of the finite groups,
where modular forms may have integer values, or more general still fractional values, called metaplectic groups~\cite{Liu:2019khw,Okada:2022kee,Ding:2022aoe,Mishra:2023cjc,Ding:2023ynd, Novichkov:2020eep,Liu:2020akv,Ding:2022nzn, Wang:2020lxk,Yao:2020zml,Behera:2021eut,Li:2021buv,Abe:2023dvr,Abe:2023qmr, Kai:2023ivp}.

In all such theories, the modular symmetry acts on the modulus field
$\tau$ in a non-linear way, and also the finite modular symmetry
is necessarily broken. $\tau$ is restricted to a fundamental domain in the upper-half complex plane which does not include zero.
However it is well known that there are three fixed points where a discrete subgroup of the modular symmetry is preserved~\cite{Novichkov:2018ovf,Novichkov:2018yse,deMedeirosVarzielas:2020kji}, namely $\tau = {\rm i} $ which preserves $Z_2^S$, $\tau = \omega = e^{2\pi{\rm i}/3}_{}$ which preserves $Z_3^{ST}$, and 
$\tau = {\rm i} \infty $ which preserves $Z_N^S$, for level $N$, where $S,T$ are the generators of the modular symmetry~\cite{Feruglio:2017spp}. At these fixed points, the Yukawa couplings may have some zero components,
which may correspond to massless charged leptons, with the charged-lepton mass hierarchy possibly resulting from small deviations 
from the fixed points~\cite{Novichkov:2021evw,Petcov:2022fjf,Petcov:2023vws,Okada:2020ukr,Wang:2021mkw,Feruglio:2021dte,Kikuchi:2022svo,Feruglio:2022koo,Feruglio:2023mii,deMedeirosVarzielas:2023crv,Kikuchi:2023jap, Abe:2023ilq}. Alternatively, the charged-lepton mass hierarchy could result from the use of so-called weighton fields \cite{King:2020qaj} which are singlet fields with non-zero modular weights which develop VEVs and provide a natural suppression mechanism for Yukawa couplings.

Since string theories are usually formulated in 10 dimensions, the simplest factorisable compactifications require three tori, which motivates bottom-up models based on three moduli fields $\tau_i$ \cite{deMedeirosVarzielas:2019cyj}
and several realistic models have been constructed along these lines \cite{King:2019vhv,King:2021fhl,deMedeirosVarzielas:2022fbw,deAnda:2023udh,deMedeirosVarzielas:2021pug, deMedeirosVarzielas:2022ihu,deMedeirosVarzielas:2023ujt}. In particular the finite fixed points $\tau = {\rm i} $  and $\tau = \omega$ seem to play a special role in modular symmetry, since they emerge from 10d supersymmetric orbifold examples~\cite{Fischer:2012qj}. 
Realistic orbifold models with three $S_4$ modular symmetries
have been constructed based on these fixed points, with two of the moduli $\tau = {\rm i} $ and $\tau = {\rm i} +2$
controlling the neutrino sector, and the third modulus $\tau = \omega$ being responsible for (diagonal) charge lepton Yukawa matrices 
\cite{deAnda:2023udh}. For the chosen orbifold $(T^2)^3/(Z_2\times Z_2)$, two of the moduli are constrained to lie at $\tau = {\rm i} $,
or equivalently $\tau = {\rm i} $ and $\tau = {\rm i} +2$, while the third modulus is not fixed by the orbifold, but was chosen to be at 
$\tau = \omega$ for phenomenological reasons, although it was observed that this choice enhanced the remnant symmetry of the 
orbifold \cite{deAnda:2023udh}. It would be interesting to see if such choices of moduli fields are stabilised at these points.

Interestingly, the minima of the effective supergravity potentials which are used to stabilise the moduli,
also seem to be situated close to the fixed points $\tau = {\rm i} $ and $\tau = \omega$.
Indeed, the most important physical implication of string theory might be the existence of extra dimensions, and the moduli are the most important particle species arising in the compactifications of extra dimensions~\cite{Cicoli:2023opf}. In this regard, modulus stabilisation is crucial for giving moduli nonzero masses and arriving at phenomenologically variable models. 
One important question is whether the minima of the potential are precisely at the fixed points $\tau = {\rm i} $ and $\tau = \omega$, or are close to these fixed points but not precisely at them. In the former case, fermion mass hierarchies could arise from the weighton fields~\cite{King:2020qaj}, while in the latter case they could arise from the deviations from the fixed points~\cite{Novichkov:2021evw} as discussed above.

One approach to modulus stabilisation is the flux compactifications, which is widely discussed in Type IIB string theory~\cite{Giddings:2001yu, Gukov:1999ya, Curio:2000sc, Ashok:2003gk, Denef:2004cf}. In the context of modular flavour symmetry, the authors in Ref.~\cite{Ishiguro:2020tmo} consider the 3-form flux in Type IIB model. They systematically analyse the stabilisation of complex structure moduli in possible configurations of flux compactifications on a $(T^2_{})^3_{}/(Z^{}_2 \times Z^{}_2)$ orbifold. The number of stabilised moduli depends on an integer $N^{\rm max}_{\rm flux}$ associated with the fluxes. The values of moduli are found to be clustered at the fixed point $\tau = \omega$ in the fundamental domain. 

Another origin of the non-trivial scalar potential is the non-perturbative effects. In Refs.~\cite{Kobayashi:2019xvz, Kobayashi:2019uyt}, the authors realise the modulus stabilisation by constructing a simple non-perturbative superpotential induced by the hidden dynamics within the framework of supergravity. In heterotic strings, there is an important non-perturbative effect called gaugino condensation~\cite{Dine:1985rz, Nilles:1982ik, Ferrara:1982qs}. Although the potential is flat in terms of the dilaton, K\"ahler and complex structure moduli at tree level, it is indeed shown that threshold corrections~\cite{Kaplunovsky:1987rp, Dixon:1990pc, Antoniadis:1991fh, Antoniadis:1992rq} or worldsheet instantons can uplift the potential and lead to non-trivial vacua~\cite{Cicoli:2013rwa}. In the presence of modular symmetries, the authors in Refs.~\cite{Font:1990nt, Gonzalo:2018guu} consider the stabilisation of K\"ahler moduli. They enumerate all possible non-perturbative contributions and derive the scalar potential. Minimising the scalar potential, they find that the anti de Sitter (AdS) vacua can generally appear at the imaginary axis and the lower boundary of the fundamental domain. They comment that no de Sitter (dS) vacuum is found in their numerical calculations. They also discuss the case where the dilaton comes into the superpotential, and argue that their results will not change if the superpotential relies on the dilaton as a sum of exponentials. The authors of Ref.~\cite{Novichkov:2022wvg} adopt the same framework. However, they find that in a special case, the VEV of $\tau$ can actually be in the interior of the fundamental domain, which is very close to the fixed point $\tau = \omega$. Still, no dS vacuum is found.

Cosmological observations imply our Universe is in a dS phase with a positive cosmological constant. If we believe the string theory is the correct ultraviolet-complete theory of particle physics and gravity, the string compactifications should yield the 4d dS cosmology. It is then interesting to investigate how to uplift the AdS vacua obtained in the simple gaugino condensation to the dS vacua. The authors of Ref.~\cite{Ishiguro:2022pde} find that non-perturbative effects and uplifting terms can lead to dS vacua around fixed points in the Type IIB theory. In Ref.~\cite{Knapp-Perez:2023nty}, the authors show that the AdS vacua can be uplifted by the matter superpotential~\cite{Lebedev:2006qq, Lebedev:2006qc}. They introduce a heavy meson field, which couples with the moduli in the K\"ahler potential and superpotential. Due to the existence of the meson field, the vacua can be uplifted to dS vacua, and the VEVs of $\tau$ could slightly deviate from the fixed points.

There are, however, still some possibilities to realise the dS vacua without introducing the matter superpotential. In Ref.~\cite{Leedom:2022zdm}, the authors investigate the modulus stabilisation within the framework of one K\"ahler modulus plus one dilaton. They first prove three no-go theorems that forbid dS vacua, which verify previous conjectures in Refs.~\cite{Font:1990nt, Gonzalo:2018guu, Novichkov:2022wvg}. In order to evade the dS no-go theorems, they further include Shenker-like effects~\cite{Shenker:1990uf} as non-perturbative corrections to the dilaton K\"ahler potential. As a result, they obtain metastable dS vacua at the fixed points $\tau = {\rm i}$ and $\omega$.

In this paper, we shall consider a modulus stabilisation mechanism which is capable of providing dS minima precisely at the fixed points
$\tau = {\rm i} $ and $\tau = \omega$, in the absence of matter fields, but taking into account the effect of the dilaton field, with non-perturbative corrections to the dilaton K\"ahler potential, along the lines of Ref.~\cite{Leedom:2022zdm},
but extended to the three-modulus case. We find that the finite fixed points can serve as dS vacua. Due to the existence of additional K\"ahler moduli, the vacuum structure becomes more diverse, and we thereby classify the different possible vacua. Conditions for these vacua to be dS vacua are distinct from those in the single-modulus case. Moreover, the dS vacua obtained at the fixed points can be the deepest under certain conditions, which is also different from Ref.~\cite{Leedom:2022zdm}. In addition, we also discuss the relation between the modulus stabilisation mechanism studied in this paper and neutrino mass models with multiple modular symmetries.

The layout of the remainder of the paper is as follows. In Sec.~\ref{sec:basic} we review the basic knowledge about modular symmetries and non-perturbative effects in the string theory, and construct the scalar potential relevant for modulus stabilisation. We study the modulus stabilisation and investigate its phenomenological implications for lepton masses and flavour mixing in Sec.~\ref{sec:stable}. We summarise our main conclusion in Sec.~\ref{sec:summary}.

\section{The modular-invariant scalar potential}\label{sec:basic}
\subsection{Modular symmetry}
\begin{figure}[t!]
    \centering
    \includegraphics[width=0.92\linewidth]{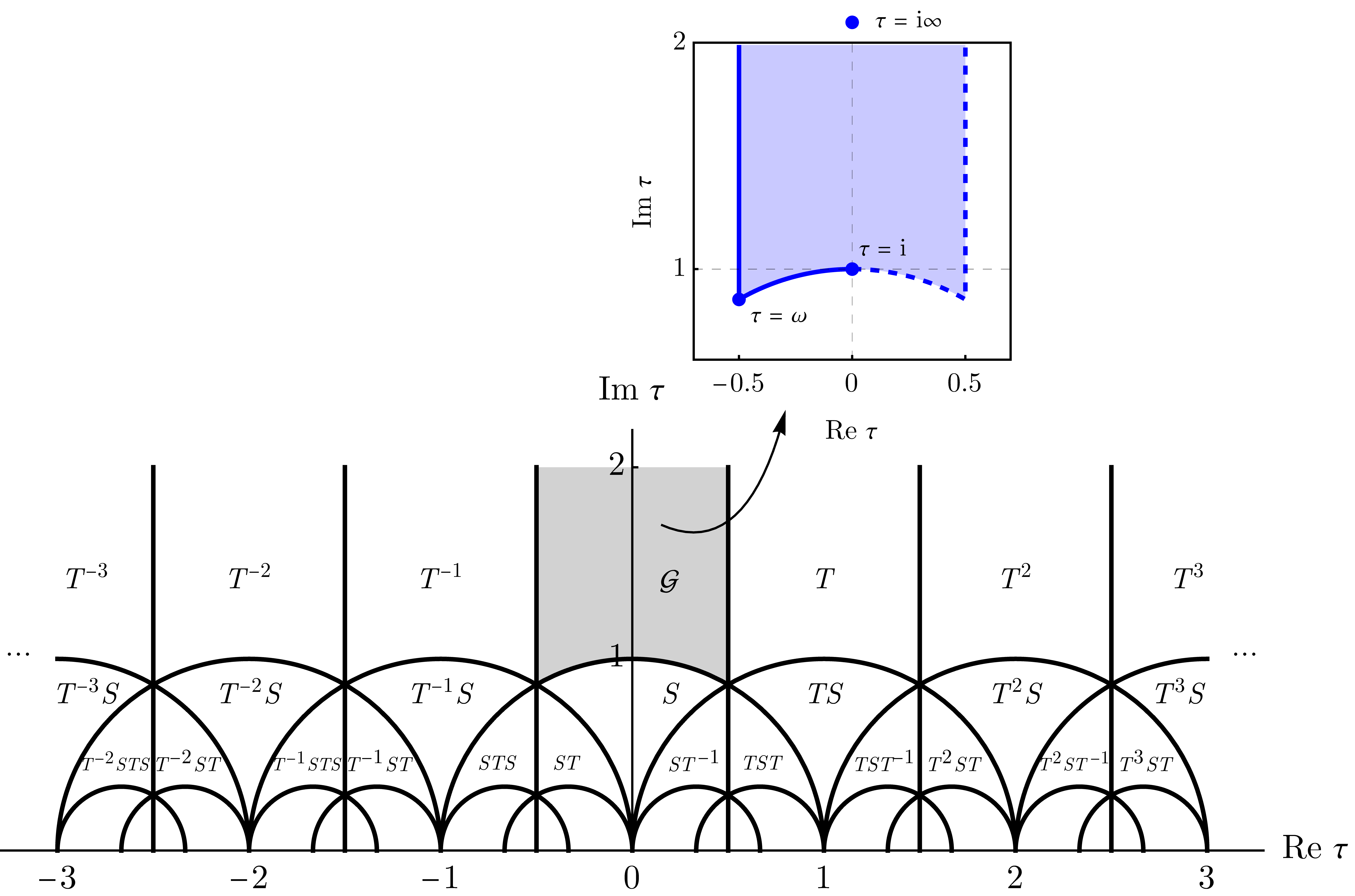}
    \caption{Fundamental domain ${\mathcal G}$ of $\overline{\Gamma}$. Acting $\overline{\Gamma}$ on $\mathcal{G}$ generates the entire upper-half complex plane with ${\rm Im}\,\tau > 0$. Three fixed points $\tau = {\rm i}$, $\omega$ and ${\rm i}\infty$ in $\mathcal{G}$ are labelled by blue dots. }
    \label{fig:GammaFD}
\end{figure}
To start with, we briefly review some basic knowledge about modular symmetries. The modular group $\overline{\Gamma}$ is isomorphic to ${\rm PSL}(2,\mathbb{Z})$ defined as~\cite{Feruglio:2017spp}
\begin{eqnarray}
\overline{\Gamma} \equiv \left\{\left(\begin{matrix}
a && b \\
c && d
\end{matrix}\right) / (\pm {\bf I}) \; \bigg| \; a, b, c, d \in \mathbb{Z}\; ,~ a d - b c = 1 \right\} \; ,
\label{eq:modgr}
\end{eqnarray}
where ${\bf I}$ is a two-dimensional unitary matrix. Under the modular group, the modulus $\tau$ and chiral supermultiplets $\chi^{(I)}_{}$ transform as 
\begin{eqnarray}
\gamma: \tau \rightarrow \dfrac{a \tau + b}{c \tau + d} \; , \quad
\chi^{(I)}_{} \rightarrow (c \tau +d )^{-k^{}_I} \rho^{}_{I} (\gamma) \chi^{(I)} _{} \; ,
\label{eq:lintran}
\end{eqnarray}
with $\gamma$ being an element of $\overline{\Gamma}$, $k^{}_I$ denoting the weight of the chiral supermultiplet and $\rho^{}_{I}(\gamma)$ representing the unitary representation matrix of $\gamma$. There are two generators $S$ and $T$ in $\overline{\Gamma}$ satisfying $S^2_{} = (ST)^3 = {\bf I}$, the matrix representations of which can be written as
\begin{eqnarray}
S = \left(\begin{matrix}
0 && 1 \\
-1 && 0
\end{matrix}\right) \; , \quad T = \left(\begin{matrix}
1 && 1 \\
0 && 1
\end{matrix}\right) \; .
\label{eq:STmatrix}
\end{eqnarray}

If we act all the elements $\gamma \in \overline{\Gamma}$ on a given point $\tau$ in the upper-half complex plane $\mathbb{C}^{}_+ = \{ \tau \in \mathbb{C}: {\rm Im}\,\tau > 0\}$, we will obtain an orbit of $\tau$. Then one can always find a minimal connected set $\mathcal{G}$, where all the orbits intersect the interior of $\mathcal{G}$ in one and only one point. The set $\mathcal{G}$ is called the fundamental domain of $\overline{\Gamma}$ defined as
\begin{eqnarray}
{\cal G} = \left\{ \tau \in \mathbb{C}^{}_{+}:  -\frac{1}{2} \leq {\rm Re}\,\tau < \frac{1}{2}, \; |\tau| > 1 \right\} \; \cup \left\{ \tau \in \mathbb{C}^{}_{+}:  -\frac{1}{2} \leq {\rm Re}\,\tau \leq 0, \; |\tau| = 1 \right\} .
\label{eq:fundo}
\end{eqnarray}
Acting $\gamma \in \overline{\Gamma}$ on $\mathcal{G}$ will generate another fundamental domain, as shown in Fig.~\ref{fig:GammaFD}.

The modular form $f(\tau)$ is a holomorphic function of $\tau$ transforming under the modular group as 
\begin{eqnarray}
f\left(\gamma\tau\right)=(c \tau+d)^{k} f(\tau) \; , \quad \gamma \in \Gamma(N) \; ,
\label{eq:modform}
\end{eqnarray}
where the level $N$ and weight $k$ are respectively positive and even integers, and $\Gamma(N)$ denote the principle congruence subgroups of $\overline{\Gamma}$. For a given $N$, the modular forms can always be decomposed into several multiplets $Y^{(k)}_{\bf r} = (f^{}_1(\tau),f^{}_2(\tau), \cdots)^{\rm T}_{}$ that transform as irreducible unitary representations of the quotient subgroups $\Gamma^{}_{N} = \overline{\Gamma}/\Gamma(N)$, namely,
\begin{eqnarray}
Y^{(k)}_{\bf r}(\gamma\tau) = (c\tau+d)^k_{}\rho^{}_{\bf r}(\gamma)Y^{(k)}_{\bf r}(\tau)\;, \quad \gamma \in \Gamma^{}_{N}\; ,
\label{eq:formtran}
\end{eqnarray}
where $\rho^{}_{\bf r}(\gamma)$ denotes the representation matrix of $\Gamma^{}_N$. $\Gamma^{}_N$ are the finite modular groups, isomorphic to non-Abelian discrete groups, e.g., $\Gamma^{}_{3} \simeq A^{}_4$, $\Gamma^{}_{4} \simeq S^{}_4$ and $\Gamma^{}_{5} \simeq A^{}_5$.

Now consider the modular-invariant supersymmetric theories. The invariance of the action ${\cal S}$ under the modular transformations requires that the K\"ahler potential $K(\tau,\overline{\tau},\chi,\overline{\chi})$ remains unchanged up to a K\"ahler transformation ${\cal K}(\tau,\overline{\tau}, \chi, \overline{\chi}) \rightarrow {\cal K}(\tau,\overline{\tau}, \chi, \overline{\chi}) +u(\tau, \chi)+u(\overline{\tau}, \overline{\chi})$ [$u(\tau, \chi)$ itself is invariant under the modular transformation], and the superpotential ${\cal W}(\tau,\chi)$ should exactly keep invariant. For the K\"ahler potential, the minimal form subject to the K\"ahler transformation is
\begin{eqnarray}
{\cal K}(\tau, \bar{\tau}, \chi, \bar{\chi})=-h \log (-{\rm i} \tau+{\rm i} \bar{\tau})+\sum_{I} \frac{\left|\chi^{(I)}_{}\right|^{2}_{}}{(-{\rm i} \tau+{\rm i} \bar{\tau})^{k_{I}}} \; , \nonumber
\label{eq:kahler}
\end{eqnarray}
where $h$ is a positive constant. The superpotential ${\cal W}(\tau,\chi)$ can be generally written as 
\begin{eqnarray}
{\cal W}(\tau, \chi)= \sum_{p}^{}\sum_{\{I^{}_{1},\dots,I^{}_{p}\}}^{} \left(Y^{}_{I^{}_1\dots I^{}_p}(\tau)\chi_{}^{(I^{}_1)}\cdots\chi_{}^{(I^{}_p)} \right)^{}_{\bf 1}\; .
\label{eq:surpoten}
\end{eqnarray}
In order for ${\cal W}(\tau, \chi)$ to be invariant under the modular transformation, the Yukawa couplings $Y^{}_{I^{}_1\dots I^{}_p}$ should take the modular forms
\begin{eqnarray}
Y^{}_{I^{}_1\dots I^{}_p}(\gamma\tau) = (c\tau+d)^{k_Y}_{}\rho^{}_{Y}(\gamma)Y^{}_{I^{}_1\dots I^{}_p}(\tau)\;, \quad \gamma \in \Gamma^{}_N \; ,
\label{eq:modformtran}
\end{eqnarray}
where $\rho^{}_Y$ denotes the representation matrix and $k^{}_{Y}$ is the weight of $Y^{}_{I^{}_1\dots I^{}_p}(\tau)$. Note that $k^{}_{Y} = k^{}_{I^{}_1} +  \cdots + k^{}_{I^{}_p}$ and $\rho^{}_{Y} \otimes \rho^{}_{I^{}_{1}}  \otimes \cdots \otimes \rho^{}_{I^{}_{p}} \ni {\bf 1}$ should be satisfied.

The modular symmetry can be extended to the framework of multiple moduli~\cite{deMedeirosVarzielas:2019cyj}. Supposing there are a series of modular groups $\overline{\Gamma}^{1}, \overline{\Gamma}^{2}, \dots , \overline{\Gamma}^{M}$ associated with different moduli $\tau^{}_1, \tau^{}_2, \dots, \tau^{}_M$, the modular transformation of each modulus field would be
\begin{eqnarray} 
&&\gamma_i: \tau_i \to \gamma_i \tau_i = \frac{a_i \tau_i + b_i}{c_i \tau_i + d_i} \; .
\label{eq:mul-tau}
\end{eqnarray}
Similar to the single-modulus case, we can obtain a set of finite modular groups $\Gamma^{i}_{N^{}_i} = \overline{\Gamma}^{i}_{}/\Gamma^{i}_{}(N^{}_i)$.  The chiral superfield $\chi_{}^{(I)}$ then transforms under the modular group $\Gamma^{1}_{N^{}_1}\times \Gamma^{2}_{N^{}_2}\times \cdots \times \Gamma_{N^{}_M}^M$ as
\begin{eqnarray}
 \chi^{(I)}_{}(\tau^{}_1, \dots,\tau^{}_M) &\to& \chi^{(I)}_{}(\gamma^{}_1\tau^{}_1, \dots, \gamma^{}_M \tau^{}_M) \nonumber\\
 &=& \prod^{}_{i=1,\dots,M} (c^{}_i\tau^{}_i + d^{}_i)^{-k^{}_{I,i}} \bigotimes^{}_{i=1,\dots,M} \rho^{}_{I,i}(\gamma^{}_i) \chi^{(I)}_{}(\tau^{}_1, \tau^{}_2, \dots,\tau^{}_M)\; ,
 \label{eq:field_transformation2}
\end{eqnarray}
where we label the elements in $\Gamma_{N^{}_i}^i$ as $\gamma^{}_i$. In addition, $k^{}_{I,i}$ and $\rho^{}_{I,i}$ are respectively the weights of  $\chi_{}^{(I)}$ and the corresponding representation matrices in $\Gamma_{N_i^{}}^{i}$, and  $\bigotimes$ represents the outer product of the representation matrices $\rho^{}_{I,1}$, $\rho^{}_{I,2}$, $\dots$, $\rho^{}_{I,M}$. 

Correspondingly, the K\"ahler potential can be rewritten as 
\begin{eqnarray}
{\mathcal K}(\tau^{}_1,\dots,\tau^{}_M,\overline{\tau}^{}_1,\dots,\overline{\tau}^{}_M, \chi, \overline{\chi}) &=& - \sum_{i=1,...,M} h^{}_i \log(-{\rm i}\tau^{}_i + {\rm i} \overline{\tau}^{}_i) \nonumber\\
&+& \sum_{I} \, \frac{\left|\chi^{(I)}_{}\right|^2_{}}{\displaystyle \prod_{i=1,...,M} (-{\rm i} \tau^{}_i + {\rm i} \overline{\tau}^{}_i)^{k^{}_{I,i}}} 
\; .
\end{eqnarray}
The superpotential ${\cal W}(\tau^{}_1, \dots, \tau^{}_M, \chi)$ becomes a modular-invariant function of all the moduli fields as well as the superfields, which takes the form
\begin{eqnarray}
{\cal W}(\tau^{}_1, \dots, \tau^{}_M, \chi) = \sum_p  \sum_{\{I^{}_1, \dots, I^{}_p\}} 
\left(Y_{I_{1},\dots, I_{p}} \chi^{(I_1)}_{} \cdots \chi^{(I_p)}_{} \right)_{\mathbf{1}} \; , 
\end{eqnarray}
Under the modular group, the modular forms $Y_{I_{1}, ..., I_{p}}$ transform as 
\begin{eqnarray} 
&& Y_{I_{1},\dots, I_{p}}(\gamma_1 \tau_1, \dots , \gamma_M \tau_M) \nonumber\\
&& \hspace{1.2cm} = \prod_{i=1, \dots,M} (c_i\tau_i + d_i)^{k_{Y,i}}
 \bigotimes_{i=1,\dots,M} \rho^{}_{Y,i}(\gamma_i) Y_{I_{1},\dots, I_{p}}(\tau_1,\dots, \tau_M) \; .
\end{eqnarray} 

Since the modular symmetries associated with different moduli are independent of each other, one modulus field obtaining its VEV will not affect the others. Once all the moduli acquire their individual VEVs, the entire modular symmetry will be spontaneously broken down. However, there are some fixed points of $\tau$, where the modular symmetry is only partially broken and we are left with residual symmetries~\cite{Novichkov:2018ovf}. There are three different fixed points in the fundamental domain $\cal{G}$ (cf. Fig.~\ref{fig:GammaFD}), namely,
\begin{itemize}
	\item $\tau^{}_{\rm C} = {\rm i}$, which is invariant under $S$ and preserves a $Z^{S}_2$ symmetry;
	\item $\tau^{}_{\rm L} = -1/2+{\rm i}\sqrt{3}/2$, which is invariant under $ST$ and preserves a $Z^{ST}_3$ symmetry;
	\item $\tau^{}_{\rm T} = {\rm i}\infty$, which is invariant under $T$ and preserves a $Z^{T}_2$ symmetry.
\end{itemize}

It is very interesting to investigate whether these special points which are fixed by residual symmetries also have dynamical origins. This is exactly the main motivation for our work.

\subsection{$\mathcal{N}=1$ supergravity theory}
We consider the ${\cal N}=1$ supergravity theory in the Abelian heterotic orbifolds, which should generally include the dilaton, the K\"ahler moduli, the complex structure moduli, gauge fields and twisted and untwisted matter fields. Here we focus on a simple scenario where only the K\"ahler moduli $\tau^{}_i$ and the dilaton field $S$ are relevant for the scalar potential.

Let us first consider the case of one K\"ahler modulus $\tau$ plus one dilaton field $S$. In the framework of supergravity theory, supersymmetry should be regarded as a local symmetry. In this case, the K\"ahler potential and the superpotential are dependent on each other via the following modular-invariant K\"ahler function
\begin{equation}
G(\tau, \overline{\tau}, S, \overline{S}) =  \kappa^2 {\cal K}(\tau, \overline{\tau},S, \overline{S}) + \log \left|\kappa^3{\cal W}(\tau, S)\right|^2_{} \; ,
\label{eq:kahler-function}
\end{equation}
{with $\kappa^2 = 8\pi/M^2_{\rm P}$ ($M_{\rm P}$ is the Planck mass).} Assuming the K\"ahler potential of $\tau$ to be the minimal form, $ {\cal K}(\tau, \overline{\tau},S, \overline{S})$ can be essentially expressed as
    \begin{equation}
   	{\cal K}(\tau, \overline{\tau},S, \overline{S}) =  {\Lambda^2_K} [K( S, \overline{S})-3 \log(2 \,{\rm Im}\,\tau) ] \; ,
    	\label{eq:kahler-potential}
    \end{equation}
where $\Lambda_K$ is a mass scale and $K( S, \overline{S})$ represents the K\"ahler potential for the dilaton.\footnote{In fact, the K\"ahler potential for the dilaton could also depend on $\tau$. Here we neglect the $\tau$-dependence for simplicity.} At tree level, we have a simple relation $K(S,\overline{S}) \propto -{\rm log}\,(S+\overline{S})$, which is related to the 4d universal gauge coupling via $g^2_4/2 =  1/\langle S + \overline{S} \rangle$ once the dilaton gets its VEV. However, if non-perturbative effects such as the Shenker-like effects are included~\cite{Shenker:1990uf}, additional corrections $\delta K(S,\overline{S})$ could be added into $K(S,\overline{S})$. We will see later that such effects play a crucial role in generating dS vacua. 
On the other hand, it is straightforward to check that ${\rm Im}\,\tau \to |c\tau + d|^{-2}_{} {\rm Im}\,\tau$ under the modular transformation, hence $e^{{\kappa^2}\cal K}_{}$ should possess a weight of six. The modular invariance of $G(\tau, \overline{\tau})$ implies that the transformation of ${\cal K}(\tau,\overline{\tau},S,\overline{S})$ under the modular group is compensated by that of ${\cal W}(\tau)$. As a result, under the modular transformation, ${\cal W}(\tau)$ should transform as 
    \begin{equation}
    	{\cal W}(\tau) \to (c \tau + d)^{-3}_{}{\cal W}(\tau) \; ,
    	\label{eq:tran-superpotential}
    \end{equation}
    indicating the superpotential possesses a weight of $-3$. In the next subsection, we will show that the superpotential satisfying Eq.~(\ref{eq:tran-superpotential}) can be induced by a non-perturbative effect---gaugino condensation.

    Once the K\"ahler potential and superpotential are known, we can construct the scalar potential $V$ as~\cite{Cremmer:1982en}
    \begin{equation}
    	V=e^{{\kappa^2}\cal K}_{}\left({\cal K}^{i \bar{j}}_{} D^{}_i {\cal W} D^{}_{\bar{j}} {\cal W}^*_{}-3 {\kappa^2}|{\cal W}|^2_{}\right) \; ,
    	\label{eq:scalar-potential}
    \end{equation}
    where $D^{}_i = \partial^{}_i + (\partial^{}_i {\cal K})$ with $\partial^{}_i$  being the first derivatives with respect to the K\"ahler moduli (which is simply $\partial/\partial \tau$ in the single-modulus case) together with the dilaton, and ${\cal K}^{i\overline{j}}_{}$ is the inverse of the K\"ahler metric ${\cal K}^{}_{i\overline{j}} = \partial^{}_i \partial^{}_{\overline{j}}{\cal K}$. The scalar potential given in Eq.~(\ref{eq:scalar-potential}) is modular-invariant, which is proved in appendix~\ref{app:A}.

    \subsection{Gaugino condensation}
    In the heterotic string constructions, gaugino condensation is a simple example that can lead to the spontaneous breakdown of supersymmetry. A gauge group $G^{}_a$ undergoing gaugino condensation will give rise to a non-perturbative superpotential of the form~\cite{Dine:1985rz, Nilles:1982ik, Ferrara:1982qs}
    \begin{equation}
       {\cal W} \sim  e^{-f_a/b_a}_{} \; ,
        \label{eq:gau-W}
    \end{equation}
    where $f_a$ is the gauge kinetic function and $b^{}_a$ is the beta function of the group $G_a$. 
    The gaugino condensation typically occurs at an energy scale $\Lambda_W \sim 10^{16}~{\rm GeV}$~\cite{Nilles:1982ik}. At the tree level, the gauge kinetic function simply takes the form $f_a = k_a S$ with $k_a$ being the level of the Kac-Moody algebra of $G_a$, which is apparently moduli-independent. However, if the orbifolds of our interest arise in ${\cal N}=2$ subsectors, threshold corrections to the gauge kinetic functions induced by integrating out heavy string states should be taken into consideration~\cite{Giddings:2001yu, Gukov:1999ya, Curio:2000sc, Ashok:2003gk, Denef:2004cf}. In the single-modulus case, the modified $f_a$ can be written as
    \begin{equation}
        f_a = k_a S + b_a {\rm log}\,\eta^6(\tau) + \cdots \; ,
        \label{eq:kinetic-f}
    \end{equation}
    where $\eta(\tau)$ is the Dedekind $\eta$ function (See appendix~\ref{app:B} for the definition). The modulus-dependent term $b_a {\rm log}\,\eta^6(\tau)$ indicates that ${\cal W}$ indeed transforms under the modular group with a weight of $-3$, and the dots denote additional contributions to threshold corrections which are also modulus-dependent but keep invariant under the modular transformation. Apart from the threshold corrections, one-loop anomaly cancellation could also lead to significant modifications to $f_a$~\cite{Derendinger:1991hq, Lust:1991yi, LopesCardoso:1991ifk, LopesCardoso:1992yd}, which however can be absorbed into $S$ by redefining the dilaton field~\cite{Kaplunovsky:1995jw}. Substituting Eq.~(\ref{eq:kinetic-f}) into Eq.~(\ref{eq:gau-W}), we arrive at the following parameterised form of ${\cal W}$
    \begin{equation}
        {\cal W}(\tau,S) = {\Lambda^3_W}\frac{\Omega(S)H(\tau)}{\eta^6(\tau)} \; ,
        \label{eq:superp-para}
    \end{equation}    
    where $\Omega(S)$ denotes a function of the dilaton field $S$, and $H(\tau)$ is a dimensionless modular-invariant function.\footnote{For the single gaugino condensation, a generic form of $\Omega(S)$ should be $\Omega(S) = v  + e^{-S/b_a}$ with $v$ being a constant.}  We can further require $H(\tau)$ to be a rational function to avoid any singularity in the fundamental domain, thus the most general form of $H(\tau)$ should be~\cite{Cvetic:1991qm}
    \begin{equation}
        H(\tau) = (j(\tau)-1728)^{m/2}j(\tau)^{n/3}{\cal P}(j(\tau)) \; ,
        \label{eq:H-def}
    \end{equation}
    with $j(\tau)$ being the modular-invariant Klein $j$ function invariant under the modular transformation defined in appendix~\ref{app:B}, $m$ and $n$ being non-negative integers and ${\cal P}$ denoting a polynomial with respect to $j(\tau)$. In the following, we take ${\cal P} = 1$ for simplicity. It is interesting to mention that $j(\omega) = 0$ and $j({\rm i}) = 12^3 = 1728$ are satisfied at the two fixed points $\tau = \omega$ and ${\rm i}$, respectively. Hence $H(\tau)$ would be vanishing at $\tau = {\rm i}$ (or $\omega$) if $m \neq 0$ (or $n \neq 0$).

    Once we  substitute the K\"ahler potential in Eq.~(\ref{eq:kahler-potential}) and superpotential in Eq.~(\ref{eq:superp-para}) into Eq.~(\ref{eq:scalar-potential}), the single-modulus scalar potential can be immediately expressed as~\cite{Leedom:2022zdm}
     \begin{equation}
        V(\tau, \overline{\tau}, S, \overline{S}) = {\Lambda_V^4}\mathcal{C}(\tau, \overline{\tau}, S, \overline{S}) \left[\mathcal{M}(\tau, \overline{\tau}) +\left(\mathcal{A}(S,\overline{S})-3\right)|H(\tau)|^2_{}\right] \; ,
        \label{eq:single-ponten}
    \end{equation}
    with
    \begin{equation}
        \begin{split}
            \mathcal{C}(\tau, \overline{\tau}, S, \overline{S}) & = \dfrac{e^{K(S,\overline{S})}_{}|\Omega(S)|^2_{}}{(2\,{\rm Im}\,\tau)^3_{}|\eta(\tau)|^{12}_{}} \; , \\
            \mathcal{M}(\tau, \overline{\tau}) & = \frac{(2\,{\rm Im}\,\tau)^2_{}}{3}\left|{\rm i}H^{\prime}_{}(\tau)+\frac{3H(\tau)}{2\pi}\widehat{G}^{}_2(\tau,\overline{\tau})\right|^2_{} \; , \\
            \mathcal{A}(S,\overline{S}) & = \frac{|\Omega^{}_S+K^{}_S\Omega|^2_{}}{K^{}_{S\overline{S}}|\Omega|^2_{}} \; ,
        \end{split}
        \label{eq:single-pon-def}
    \end{equation}
    where we have {$\Lambda_V = (\kappa^2 \Lambda_W^6)^{1/4}$} and $K^{S\overline{S}} = (K_{S\overline{S}})^{-1}_{}$, and the subscripts $S$ and $\overline{S}$ represent the first derivatives with respect to $S$ and $\overline{S}$, respectively. Moreover, $\widehat{G}_2$ is the non-holomorphic Eisenstein function of weight two defined as
    \begin{equation}
        \widehat{G}_2(\tau,\overline{\tau}) = G_2(\tau)-\frac{\pi}{{\rm Im}\,\tau} \; ,
        \label{eq:nh-eisenstein}
    \end{equation}
    where the Eisenstein series $G_2$ is a holomorphic counterpart of $\widehat{G}_2(\tau,\overline{\tau})$, and can be related to the Dedekind $\eta$ function via
    \begin{equation}
    \frac{\eta^\prime_{} (\tau)}{\eta(\tau)} = \frac{\rm i}{4\pi} G_2(\tau) \; .
    \label{eq:hol-eisen}
    \end{equation}

\section{Modulus stabilisation} \label{sec:stable}
    Before going into the details of minimising the scalar potential, we can first gain some general insights without specifying the form of the scalar potential. One salient feature of the scalar potential is that it diverges in the limit ${\rm Im}\,\tau \to \infty$. Hence the fixed point $\tau \to {\rm i}\infty$ seems not to be the vacuum. The finite fixed points, however, are able to be the minima of the scalar potential. In fact,  since $V$ is a zero-weight modular form, $\partial V/\partial \tau$ must be a modular form with weight two (See appendix~\ref{app:A} for proof). Then if we consider the modular transformation of $\partial V/\partial \tau$ under the generator $S$ at $\tau = {\rm i}$, we will arrive at
	\begin{equation}
			\left.(\partial V/\partial \tau)\right|^{}_{\tau = {\rm i}} = (-{\rm i})^2_{} \left.(\partial V/\partial \tau)\right|^{}_{\tau = {\rm i}} = -\left. (\partial V/\partial \tau)\right|^{}_{\tau = {\rm i}} \; .
			\label{eq:deriv-i}
	\end{equation}
    Similarly, if we consider the modular transformation of $\partial V/\partial \tau$ under the generator $ST$ at $\tau = \omega$, we will obtain
	\begin{equation}
	\left.(\partial V/\partial \tau)\right|^{}_{\tau = \omega} = (-\omega-1)^2_{} \left.(\partial V/\partial \tau)\right|^{}_{\tau =\omega} = \omega \left.(\partial V/\partial \tau)\right|^{}_{\tau = \omega} \; .
	\label{eq:deriv-omega}
\end{equation}
Eqs.~(\ref{eq:deriv-i}) and (\ref{eq:deriv-omega}) tell us $\partial V/\partial \tau$ has to be zero at $\tau = {\rm i}$ and $\omega$. Therefore the finite fixed points should be the extrema of the scalar potential in the K\"ahler moduli space. However, identifying whether they are exactly the minima requires an in-depth analysis of certain scalar potentials.

Another important aspect is that the dilaton sector also plays a crucial role in modulus stabilisation. The effects of the dilaton are mainly two-fold. On the one hand, a positive $\cal{A}(S,\overline{S})$ term in the superpotential could uplift the minima to dS vacua. On the other hand, stringy corrections to the K\"ahler potential $K(S,\overline{S})$ are required to stabilise the potential in the dilaton sector. In fact, the scalar potential should satisfy $\partial V/\partial S =0$ at the minima. Hence we arrive at
\begin{equation}
    \frac{\partial V}{\partial S} = \dfrac{ {\Lambda_V^4} e^{K}_{}}{(2\,{\rm Im}\,\tau)^3_{}|\eta(\tau)|^{12}_{}}(\Omega^{}_S+K^{}_S\Omega){\mathcal Q} = 0 \; ,
    \label{eq:stable-S-eqn}
\end{equation}
where  
\begin{align}
    {\mathcal Q} = & e^{-2{\rm i}\sigma}|H|^2\left[ (\Omega^{}_S+K^{}_S\Omega)\left(\frac{K^{}_S}{K^{}_{S\overline{S}}} - \frac{K^{}_{SS\overline{S}}}  {K^2_{S\overline{S}}} \right) + \frac{\Omega^{}_{S\overline{S}}}{K^{}_{S\overline{S}}} + \frac{\Omega K^{}_{SS}}{K^{}_{S\overline{S}}} + \frac{\Omega^{}_S K^{}_S}{K^{}_{S\overline{S}}} \right] \nonumber \\
    & + \overline{\Omega}\left(\mathcal{M}-2|H|^2\right) \; ,
    \label{eq:def-Q}
    \end{align}
with $\sigma$ being the phase angle of $\Omega^{}_S+K^{}_S\Omega$. Then we can immediately gain the following two possibilities of the necessary conditions for $S$ to be stabilised
\begin{equation}
    \begin{split}
    {\it Condition~A:} \quad &\Omega^{}_S+K^{}_S\Omega = 0 \; ; \\
    {\it Condition~B:} \quad &\Omega^{}_S+K^{}_S\Omega \neq 0 \; , \quad {\cal Q} = 0 \; .
    \end{split}
\end{equation}
Indeed, {\it Condition A} corresponds to the case where ${\mathcal A}(S, \overline{S})$ is vanishing, i.e., the scalar potential can be written as a factorised form of the dilaton and K\"ahler moduli sector. In Ref.~\cite{Leedom:2022zdm}, the authors have proved three no-go theorems regarding the dS vacua under {\it Condition A}, indicating that {\it Condition A} can never lead to dS vacua, i.e., the dependence of the scalar potential on the dilation and K\"ahler moduli should not be factorised. Hence one must switch to {\it Condition B} in order to obtain dS vacua.

{Nevertheless, even if it is possible for the extrema that satisfy {\it Condition B} to be the dS vacua, such vacua may still be unstable in the dilaton sector. Indeed, one can prove that if only the tree-level K\"ahler potential for the dilaton $K(S,\overline{S}) \propto -{\rm log}\,(S + \overline{S})$ is included, $\tau = {\rm i}$ and $\omega$ could never be the dS vacua no matter which form $\Omega(S)$ takes since the extrema are unstable in the dilaton sector~\cite{Leedom:2022zdm}. In order to evade the dS no-go theorems, one should go beyond the minimal K\"ahler potential of $S$. It is found in Ref.~\cite{Leedom:2022zdm} that non-perturbative Shenker-like effects can result in non-trivial corrections to $K(S,\overline{S})$, rendering the dilaton sector metastable at the fixed points $\tau = {\rm i}$ and $\omega$. Different from the gaugino condensation which has a generic strength $\delta {\cal L} = e^{-1/g^2_s}_{}$ with $g^{}_s$ being the string coupling constant, Shenker-like effects are inherently stringy effects which lead to modifications of ${\mathcal O}(e^{-1/g^{}_s})$. In the rest of this paper, we first explore how ${\cal A}(S,\overline{S})$ can modify the modulus stabilisation assuming the dilaton has been stabilised, and then we show concrete examples of the Shenker-like effects that can generate non-trivial $\calAS$ in appendix~\ref{app:D}.}
\subsection{Minimising the single-modulus scalar potential}\label{susec:single-stabilisation}
{For the convenience of the readers, we first briefly review the minimisation of the scalar potential with a single modulus, which has been widely studied under both {\it Condition A} and {\it Condition B} in the previous literature~\cite{Font:1990nt, Cvetic:1991qm, Gonzalo:2018guu, Novichkov:2022wvg}.} In Refs.~\cite{Cvetic:1991qm, Gonzalo:2018guu}, assuming ${\mathcal A}(S, \overline{S}) = 0$, the authors analyse different scalar potentials by varying the indices $m$ and $n$ in Eq.~(\ref{eq:H-def}). They have numerically searched the minima of the scalar potentials and concluded that the vacua should appear either on the lower boundary of the fundamental domain or on the imaginary axis of $\tau$. The authors in Ref.~\cite{Novichkov:2022wvg} find a special case where $m \neq 0$ and $n = 0$ can lead to global minima very close to but not precisely at the fixed point $\tau = \omega$. In summary, there are four different types of {AdS} vacua depending on the choices of $m$ and $n$ if {\it Condition A} is satisfied:
\begin{itemize}
    \item $m=0$ and $n=0$: The vacuum $\langle \tau \rangle = 1.235{\rm i}$;
    \item $m = 0$ and $n \neq 0$: The vacuum $\langle \tau \rangle = {\rm i}$;
    \item $m \neq 0$ and $n = 0$: The vacua are close to but not precisely at $\tau = \omega$;
    \item $m \neq 0$ and $n \neq 0$: The vacua are located at the lower boundary of the fundamental domain.
\end{itemize}

A detailed analysis of minimising the one-modulus scalar potential {under {\it{Condition B}}} can be found in Ref.~\cite{Leedom:2022zdm}. {In this case, the finite fixed points can be dS or Minkowski vacua. Types of these vacua dramatically depend on the values of ${\cal A}(S,\overline{S})$, which can be analysed by calculating the Hessian matrices (See appendix~\ref{app:C})}. The main conclusions are collected as follows:
\begin{itemize}
    \item $m=0$ and $n=0$: $\tau = \omega$ is always the dS vacuum, while $\tau = {\rm i}$ can be the dS vacuum if $3<{\cal A}(S,\overline{S})<3.5964$ is satisfied;
    \item $m > 1$, $n = 0$: $\tau = \omega$ is always a dS vacuum if ${\cal A}(S,\overline{S})>3$, and $\tau = {\rm i}$ is a Minkowski vacuum;
    \item $m = 0$, $n >1$: $\tau = {\rm i}$ is a dS vacuum within a window of ${\cal A}(S,\overline{S})$ which increases with $n$, and $\tau = \omega$ is a Minkowski vacuum.
    \item $m = 1$ or $n=1$: $\tau ={\rm i}$ or $\tau = \omega$ could be the minimum in terms of the K\"ahler modulus, but it is actually unstable in the dilaton sector;
    \item $m >1$, $n >1$: Both $\tau = {\rm i}$ and $\tau = \omega$ are Minkowski vacua when ${\cal A}(S,\overline{S})>3$.
\end{itemize}

\begin{figure}[t!]
    \centering
    \includegraphics[width=1\linewidth]{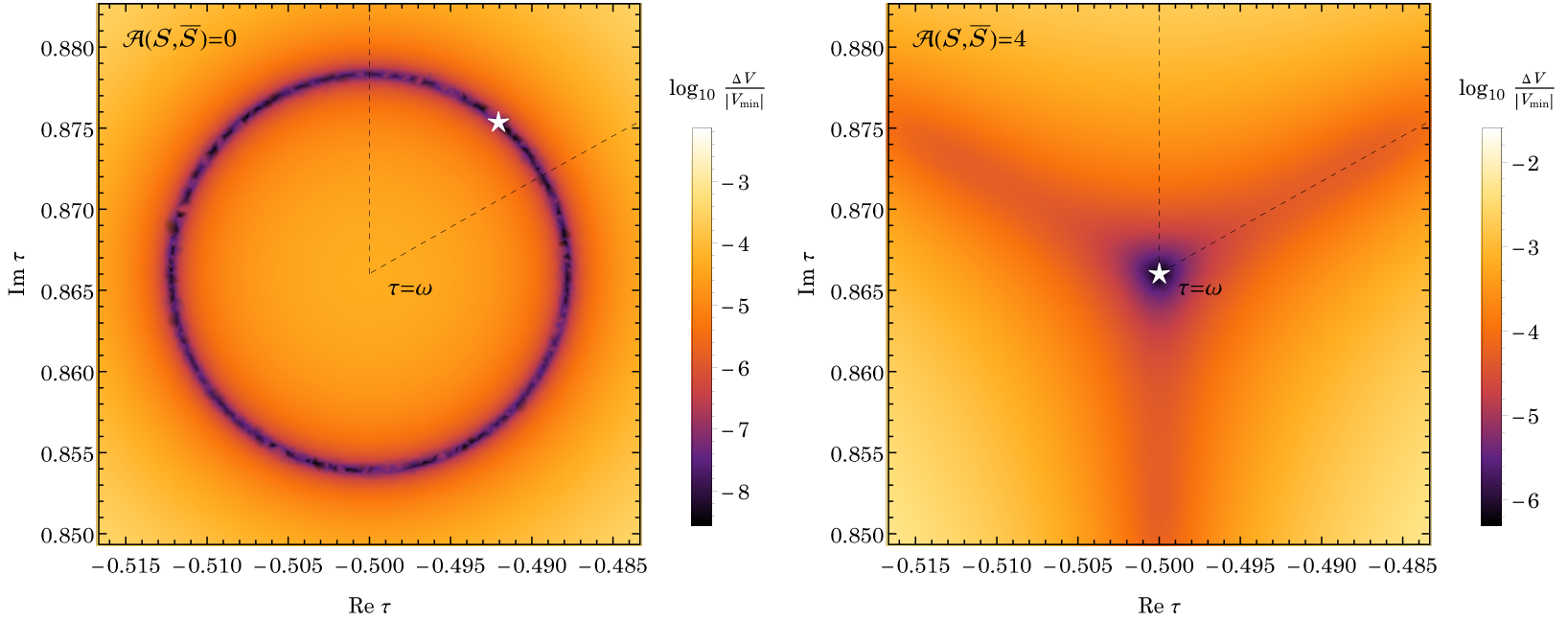}
    \caption{Density plots of the distribution of ${\rm log}_{10}(\Delta V/|V^{}_{\rm min}|)$  with $\Delta V = V - V^{}_{\rm min}$ in the vicinity of $\tau = \omega$. We take $m = 2$ and $n =0$ for instance. In the left panel, we choose ${\cal A}(S,\overline{S}) = 0$, while ${\cal A}(S,\overline{S})$ is fixed to be four in the right panel. We use white stars to label the vacua. The dashed lines correspond to the boundaries of the fundamental domain.}
    \label{fig:omega}
\end{figure}

 As can be seen above, the inclusion of dilaton effects will not only uplift the vacua to dS/Minkowski vacua, but also shift the VEVs of $\tau$ towards the fixed points. For illustration, we consider a specific case with $m =2$ and $n = 0$. In Fig.~\ref{fig:omega}, we exhibit the distribution of ${\rm log}_{10}(\Delta V/|V^{}_{\rm min}|)$ with $\Delta V$ defined as the difference between $V$ and its minimal value $V_{\rm min}$ in the vicinity of $\tau = \omega $ under the assumptions ${\cal A}(S,\overline{S}) = 0$ and ${\cal A}(S,\overline{S}) = 4$, respectively. In the case where ${\cal A}(S,\overline{S}) = 0$,  the fixed point $\tau = \omega$ turns out to be a local maximum, and the global AdS vacuum appears at $\tau = -0.492 + 0.875{\rm i}$ with $V_{\rm min} = -2.48 \times 10^7_{} {\Lambda^4_V}$, which is consistent with the result in Ref.~\cite{Novichkov:2022wvg}. However, if ${\cal A}(S,\overline{S}) > 3$, e.g., ${\cal A}(S,\overline{S}) = 4$, we obtain a vacuum precisely at $\tau = \omega$, where the value of $V$ is found to be $V_{\rm min} = 8.29 \times 10^6_{}{\Lambda_V^4}$, indicating $\tau = \omega$ is indeed a dS vacuum. 

 \begin{figure}[t!]
    \centering
    \includegraphics[width=1\linewidth]{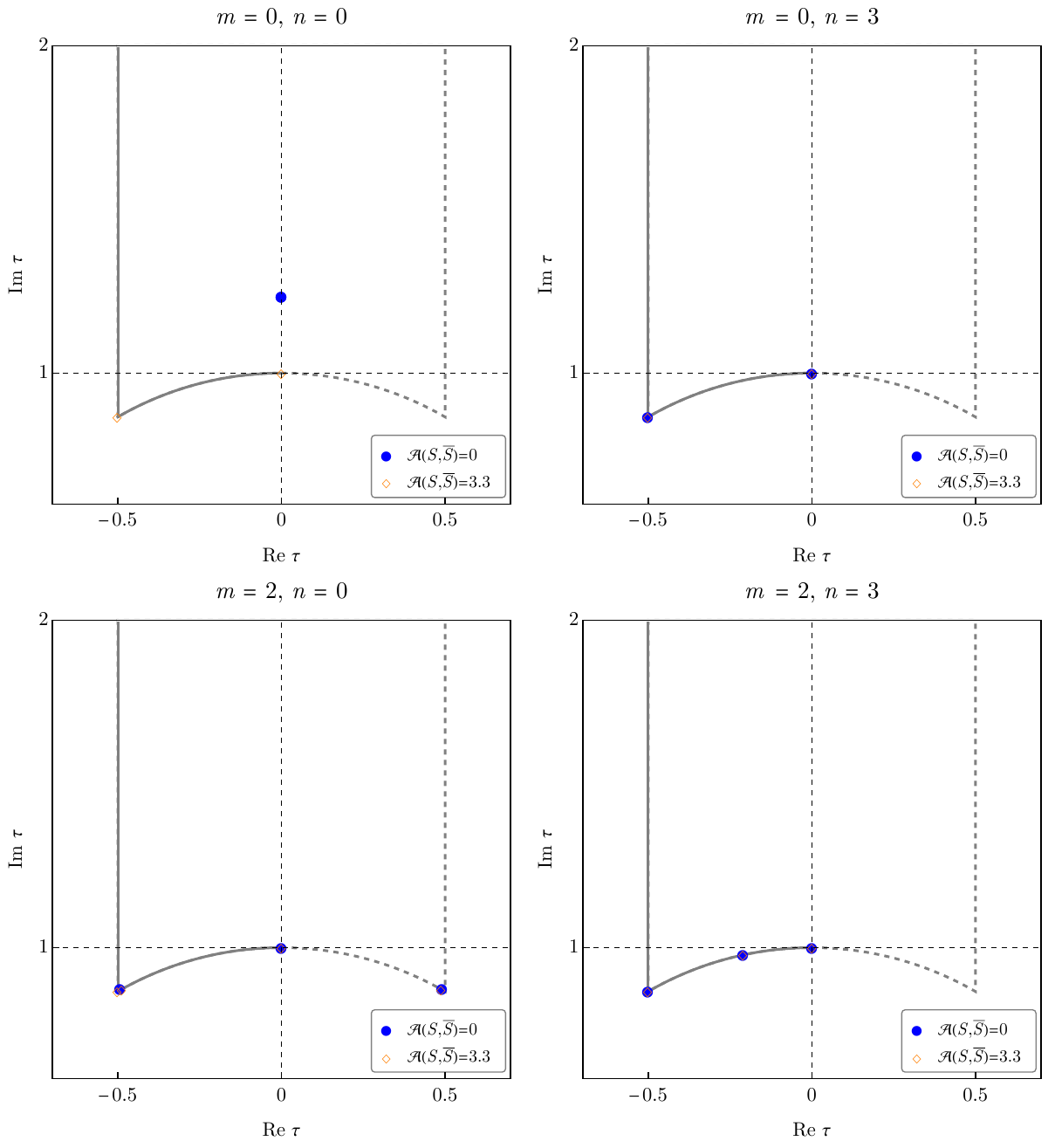}
    \caption{Complete sets of the minima of the single-modulus scalar potential in the fundamental domain, where the blue dots and orange diamonds represent the vacua obtained with vanishing  ${\cal A}(S,\overline{S})$ and non-vanishing ${\cal A}(S,\overline{S})$ [${\cal A}(S,\overline{S}) = 3.3$], respectively. Different values of $m$ and $n$ are taken into consideration.}
    \label{fig:dSI}
\end{figure}

{Moreover, the presence of the dilaton term in the superpotential may also transition the fixed points into global minima. This can be understood by considering the expression of the scalar potential in Eq.~(\ref{eq:single-ponten}). Since both ${\cal M}(\tau,\overline{\tau})$ and $|H(\tau)|^2$ are non-negative, the scalar potential would be semi-positive definite if ${\cal A}(S,\overline{S}) \geq 3$. ${\cal M}(\tau,\overline{\tau})$ depends on $H(\tau)$ and $H^\prime(\tau)$, both of which could be vanishing at $\tau = {\rm i}$ or $\omega$ when $m > 1$ or $n >1$, hence there would be at least one finite fixed point corresponding to the global minimum of the scalar potential.  }

{In order to verify that finite fixed points can truly be the global vacua, we also implement a numerical approach. In specific, we initiate our analysis by randomly generating starting points in the fundamental domain. Subsequently, we employ the gradient descent technique to meticulously search for the local minima. The results are shown in Fig.~\ref{fig:dSI}, where the blue dots and orange diamonds represent the complete sets of local minima obtained with vanishing ${\cal A}(S,\overline{S})$ and non-vanishing ${\cal A}(S,\overline{S})$ [${\cal A}(S,\overline{S}) = 3.3$], respectively. By including the dilaton effects, both $\tau = {\rm i}$ and $\omega$ can be the vacua for certain ranges of ${\cal A}(S,\overline{S})$. It is interesting to point out that there are additional vacua inside the fundamental domain even if ${\cal A}(S,\overline{S}) > 3$. For example, when $m = 2$ and $n = 0$, we have another dS vacuum at $\tau = - 0.489 + 0.872{\rm i}$, which is very close to $\tau = \omega$. However, this vacuum is not the global one as $V|_{\tau = - 0.489 + 0.872{\rm i}} = 8.29 \times 10^6\Lambda^4_V$ while $V|_{\tau = {\rm i}} = 0$. In addition, when $m =2$ and $n = 3$, we can find an additional vacuum on the lower boundary of the fundamental domain with $\tau = - 0.211 + 0.978{\rm i}$, which is again not the deepest as both $\tau ={\rm i}$ and $\omega$ turn out to be Minkowski vacua.}

\subsection{Modulus stabilisation in the three-modulus framework}
Since the compactification of 10d heterotic string theory will generally lead to three moduli, associated with three 2d tori,\footnote{It should be mentioned that here we focus on the scenario where the extra 6d space can be factorised into three $T^2$ tori. However, non-factorisable toroidal manifolds~\cite{Faraggi:2006bs, Forste:2006wq, Kimura:2007ey, Blumenhagen:2004di} can have different geometries from factorisable ones since the number of fixed tori could be less. Consequently, the moduli are not separable and may be incorporated into some larger symmetry groups, e.g., the Siegel modular group~\cite{Mayr:1995rx,Baur:2020yjl,Ishiguro:2021ccl,Ding:2020zxw,Ding:2024xhz}. The scenario for modulus stabilisation in the non-separable case could be different, which is beyond the scope of the present paper.} we should extend the single modulus stabilisation into this more complete scenario, and explore how the non-perturbative effects can give a dynamical explanation of the VEVs of moduli with multiple modular symmetries. 

In the three-modulus case, the modular-invariant function $H(\tau)$ in the superpotential should be replaced by a more general form
    \begin{equation}
    \mathcal{H}(\tau^{}_1,\tau^{}_2,\tau^{}_3) = \sum^{}_{\substack{ m_1,m_2,m_3 \\ n_1,n_2,n_3}} H^{(m_1,n_1)}_{}(\tau^{}_1) H^{(m_2,n_2)}_{}(\tau^{}_2)H^{(m_3,n_3)}_{}(\tau^{}_3) \; ,
    \label{eq:H-multiple}
    \end{equation}
    where $H^{(m_i,n_i)}_{} = (j(\tau)-1728)^{m_{i}/2}_{} j(\tau)^{n_{i}/3}_{}$ for $i = 1,2,3$. {Given the infinite number of modular-invariant ${\cal H}(\tau_1,\tau_2, \tau_3)$, it is difficult to investigate the modulus stabilisation for all ${\cal H}(\tau_1,\tau_2, \tau_3)$ in a systematic way. Instead, we try to find the minimal superpotential that can lead to global dS vacua at the fixed points.} One may notice that the simplest ${\cal H}(\tau^{}_1, \tau^{}_2, \tau^{}_3)$ should be a factorised form $H^{(m_1,n_1)}_{}(\tau^{}_1) H^{(m_2,n_2)}_{}(\tau^{}_2) H^{(m_3,n_3)}_{}(\tau^{}_3)$,\footnote{The factorised form of the superpotential was considered in ref.~\cite{Cvetic:1991qm}, where $H^{(m_i,n_i)}(\tau_i)$ for all the moduli take the same form.} which, however would become zero as long as one $ H^{(m_i,n_i)}_{}(\tau^{}_i)$ is vanishing at the fixed points. Consequently, this scenario will essentially lead to Minkowski vacua at the fixed points. Instead, we consider ${\cal H}(\tau^{}_1, \tau^{}_2, \tau^{}_3)$ as the summation of three different $H^{(m_i,n_i)}_{}(\tau^{}_i)$, namely,\footnote{It seems more natural to expect a factorised form for ${\cal H}(\tau^{}_1, \tau^{}_2, \tau^{}_3)$, since the loop-level corrections from each torus contribute to the superpotential as exponential forms, as can be seen in Eq.~(\ref{eq:gau-W}). However, ${\cal H}(\tau^{}_1, \tau^{}_2, \tau^{}_3)$ in Eq.~(\ref{eq:H-fun}) may still be realised, by, e.g., introducing multiple dilatons, each of which is associated with one torus.}
    \begin{equation}
        {\cal H}(\tau^{}_1, \tau^{}_2, \tau^{}_3) = H^{(m_1,n_1)}_{}(\tau^{}_1) + H^{(m_2,n_2)}_{}(\tau^{}_2) + H^{(m_3,n_3)}_{}(\tau^{}_3) \; .
        \label{eq:H-fun}
    \end{equation}
    Then ${\cal H}(\tau^{}_1, \tau^{}_2, \tau^{}_3)$ would be nonzero as long as at least one of $H^{(m_i,n_i)}_{}(\tau^{}_i)$ is non-vanishing, making the realisation of dS vacua more likely. As a result, the K\"ahler potential and superpotential can be respectively rewritten as
    \begin{eqnarray}
            {\cal K}(\tau^{}_i,\overline{\tau}^{}_i,S, \overline{S}) &=& {\Lambda^2_K} \{K( S, \overline{S})-\log[(2 \,{\rm Im}\,\tau^{}_1)(2 \,{\rm Im}\,\tau^{}_2)(2 \,{\rm Im}\,\tau^{}_3)] \} \; , \label{eq:3-Kahler}\\
            {\cal W}(\tau^{}_i,S) &=& \frac{{\Lambda^3_W}\Omega(S)[H^{(m_1,n_1)}_{}(\tau^{}_1) + H^{(m_2,n_2)}_{}(\tau^{}_2) + H^{(m_3,n_3)}_{}(\tau^{}_3)]}{\eta^2_{}(\tau^{}_1)\eta^2_{}(\tau^{}_2)\eta^2_{}(\tau^{}_3)} \label{eq:3-super}\; ,
    \end{eqnarray}
    where the variables $\tau^{}_i$ go through $\{\tau^{}_1, \tau^{}_2, \tau^{}_3\}$. The scalar potential in this scenario turns out to be
    \begin{equation}
        V(\tau^{}_i, \overline{\tau^{}_i},S, \overline{S}) = {\Lambda^4_V}\widetilde{\mathcal{C}}(\tau^{}_i, \overline{\tau^{}_i},S, \overline{S}) \left\{ \widetilde{\mathcal{M}}(\tau^{}_i, \overline{\tau}^{}_i) +\left[\mathcal{A}(S,\overline{S})-3\right]|\mathcal{H}(\tau^{}_i)|^2_{}\right\} \; ,
        \label{eq:triple-ponten}
    \end{equation}
    with
    \begin{equation}
        \begin{split}
            \widetilde{\mathcal{C}}(\tau^{}_i, \overline{\tau^{}_i},S, \overline{S}) & = \prod_{i=1}^3 \dfrac{ e^{K(S,\overline{S})}_{}|\Omega(S)|^2_{}}{(2\,{\rm Im}\,\tau^{}_i)|\eta(\tau^{}_i)|^4_{}} \; , \\
            \widetilde{\mathcal{M}}(\tau^{}_i, \overline{\tau}^{}_i) & = \sum^3_{i=1} (2\,{\rm Im}\,\tau^{}_i)^2_{}\left|{\rm i} \frac{\partial H^{(m_i,n_i)}_{}(\tau^{}_i)}{\partial \tau^{}_i}+\sum_{j=1}^3\frac{H^{(m_j,n_j)}_{}(\tau^{}_j)}{2\pi}\widehat{G}^{}_2(\tau^{}_i,\overline{\tau}^{}_i)\right|^2_{} \; .
        \end{split}
        \label{eq:pon-def}
    \end{equation}
    One can observe that apart from $\tau^{}_{1,2,3}$ and $\mathcal{A}(S,\overline{S})$, there are six  additional parameters that can affect the minima of the scalar potential, namely, $m^{}_{1,2,3}$ and $n^{}_{1,2,3}$. In the following, we discuss the minimisation of the scalar potential given in Eq.~(\ref{eq:triple-ponten}), mainly focusing attention on the finite fixed points ${\rm i}$ and $\omega$. In order to identify whether they are indeed the minima of the potential, we again calculate the Hessian matrices at the fixed points and make them positive-definite. We also thoroughly search the minima of $V$ in the entire fundamental domain for different $m^{}_{1,2,3}$, $n^{}_{1,2,3}$ and $\mathcal{A}(S,\overline{S})$ using the gradient descent approach, which could help us identify whether the fixed points can be the global minima of the scalar potential. 

The second derivatives of $V$ in terms of K\"ahler moduli are expressed as
\begin{equation}
    \begin{split}
    \frac{\partial^2_{} V}{{\Lambda^4_V}\partial \tau^2_i} &= \frac{\partial^2_{} \widetilde{\mathcal{C}}}{\partial \tau^2_i}\left[ \widetilde{\mathcal{M}} + (\mathcal{A}-3) |\mathcal{H}|^2_{}  \right] +\widetilde{\mathcal{C}} \left[  \frac{\partial^2_{} \widetilde{\mathcal{M}}}{\partial \tau^2_i} + (\mathcal{A}-3)\mathcal{H}^*_{} \frac{\partial^2_{} H^{(m^{}_i,n^{}_i)}_{}}{\partial \tau^{2}_i}\right] \; , \\
    \frac{\partial^2_{} V}{{\Lambda^4_V}\partial \tau^{}_i \partial \overline {\tau}^{}_j} & = \frac{\partial^2 \widetilde{\mathcal{C}}}{\partial \tau^{}_i \partial \overline{\tau}^{}_j} [\widetilde{\mathcal{M}} + (\mathcal{A}-3)|\mathcal{H}|^2_{}]  + \widetilde{\mathcal{C}} \left[ \frac{\partial^2_{}\widetilde{\mathcal{M}}}{\partial \tau^{}_i \partial \overline{\tau}^{}_j}  + (\mathcal{A} -3) \left|\frac{\partial H^{(m_i,n_i)}_{}}{\partial \tau^{}_i } \right|^2_{}
    \right] \; ,
    \end{split}
  \label{eq:deri-V}
\end{equation}
with
    \begin{align}
        \frac{\partial^2 \widetilde{\mathcal{C}}}{\partial \tau^{}_i \partial \tau^{}_j} & = -{\rm i} \delta^{}_{ij} \widetilde{\mathcal{C}} \frac{\partial}{\partial \tau^{}_i } \frac{\widehat{G}^{}_2(\tau^{}_i,\overline{\tau}^{}_i)}{2\pi}  \; , \nonumber \\
        \frac{\partial^2 \widetilde{\mathcal{C}}}{\partial \tau^{}_i \partial \overline{\tau}^{}_j} & = {\rm i}\delta^{}_{ij}\widetilde{\mathcal{C}} \frac{\partial}{\partial \tau^{}_i } \frac{[\widehat{G}^{}_2(\tau^{}_j,\overline{\tau}^{}_j)]^*_{}}{2\pi}   \; , \nonumber \\
        \frac{\partial^2 \widetilde{\mathcal{M}}}{\partial \tau^{}_i \partial \tau^{}_j} & = \delta^{}_{ij} (2\,{\rm Im}\,\tau^{}_i)^2_{}\left[  {\rm i} \frac{\partial^2_{} H^{(m_i,n_i)}_{}(\tau^{}_i)}{\partial \tau^{2}_i} + \frac{\mathcal{H}(\tau^{}_i)}{2\pi} \frac{\partial \widehat{G}^{}_2(\tau^{}_i,\overline{\tau}^{}_i )}{\partial \tau^{}_i} \right] \frac{\mathcal{H}^*_{}(\tau^{}_i)}{\pi}\frac{\partial \widehat{G}^{*}_2(\tau^{}_i,\overline{\tau}^{}_i )}{\partial \tau^{}_i} \; , \nonumber \\      
        \frac{\partial^2 \widetilde{\mathcal{M}}}{\partial \tau^{}_i \partial \overline{\tau}^{}_j} & = \delta^{}_{ij} (2\,{\rm Im}\,\tau^{}_i)^2_{} \left[    \left| {\rm i} \frac{\partial^2_{} H^{(m_i,n_i)}_{}(\tau^{}_i)}{\partial \tau^{2}_i} + \frac{\mathcal{H}(\tau^{}_i)}{2\pi} \frac{\partial \widehat{G}^{}_2(\tau^{}_i,\overline{\tau}^{}_i )}{\partial \tau^{}_i} \right|^2_{} + \left| \frac{\mathcal{H}(\tau^{}_i)}{2\pi} \frac{\partial{\widehat{G}^{}_2(\tau^{}_i,\overline{\tau}^{}_i )}}{\partial \overline{\tau}^{}_i} \right|^2_{} \right] \; ,
            \label{eq:deri}
    \end{align}
where all the derivatives above are calculated at $\tau = {\rm i}$ or $\omega$. 

According to the choices of $(m^{}_i,n_i)$, we have the following three distinct classes:
\begin{itemize}
    \item {\bf Class A---}$\bm{(m^{}_1,n^{}_1)=(m^{}_2,n^{}_2) = (m^{}_3,n^{}_3)}.$ This is an exactly symmetric class, where three moduli parameters can be exchanged. It is natural to expect the global minima should appear at $\tau^{}_1 = \tau^{}_2 = \tau^{}_3$. Then one can find that $\widetilde{\mathcal C}(\tau^{}_i,\overline{  \tau}^{}_i, S, \overline{S})$ and $\widetilde{\mathcal M}(\tau^{}_i, {\overline{\tau}^{}_i})$ in Eq.~(\ref{eq:pon-def}) reduce to the single-modulus case, and thus the results for the minima are the same as those obtained in the single-modulus case.
    \item {\bf Class B---}$\bm{(m^{}_1,n^{}_1)=(m^{}_2,n^{}_2) \neq (m^{}_3,n^{}_3)}.$ In this case, we can freely exchange $\tau^{}_1$ and $\tau^{}_2$ without affecting the value of the scalar potential, hence it is effectively a two-modular case, where only two K\"ahler moduli $\tau^{}_1$ (or $\tau^{}_2$) and $\tau^{}_3$ are independent. 
    \item {\bf Class C---}$\bm{(m^{}_1,n^{}_1) \neq (m^{}_2,n^{}_2) \neq (m^{}_3,n^{}_3)}.$ This class becomes more complicated since there is no symmetry among the three moduli parameters. In this class we should consider all three moduli as free parameters.
\end{itemize}

We first focus on {\bf Class B}. In this class, the number of independent real variables is reduced to four, indicating that the Hessian matrices should be four-dimensional. On the other hand, Eq.~(\ref{eq:deri}) tells us all the mixed second derivatives in terms of different moduli are vanishing. Moreover, the imaginary parts of $\partial^2_{} V/(\partial \tau^{}_i \partial \tau^{}_j)$ are also zero at the finite fixed points. Hence we arrive at the following diagonal Hessian matrices
\begin{equation}
    {\mathbf H} =
    \left( 
    \begin{matrix}
        \dfrac{\partial^2_{} V}{\partial s^{2}_1} & 0 & 0 & 0 \\
        0 & \dfrac{\partial^2_{} V}{\partial t^{2}_1} & 0 & 0 \\
        0 & 0 & \dfrac{\partial^2_{} V}{\partial s^{2}_3} & 0 \\
        0 & 0 & 0  & \dfrac{\partial^2_{} V}{\partial t^{2}_3} 
    \end{matrix}
    \right) \; .
    \label{eq:Hessian}
\end{equation}
Therefore, in order for finite fixed points to be the minima of the scalar potential, we should require each element in Eq.~(\ref{eq:Hessian}) to be positive. 

Since effectively we have two independent modulus parameters $\tau^{}_1$ and $\tau^{}_3$, there are twelve kinds of arrangements of the indices $(m^{}_i, n^{}_i)$ depending on whether they are zero or not, including
\begin{itemize}
    \item $(m^{}_1, n^{}_1) = (0,0) $, $(m^{}_3, n^{}_3) = (0,\underline{n}^{}_3) $; 
    \item $(m^{}_1, n^{}_1) = (0,0) $, $(m^{}_3, n^{}_3) = (\underline{m}^{}_3,0) $; 
    \item $(m^{}_1, n^{}_1) = (0,0) $, $(m^{}_3, n^{}_3) = (\underline{m}^{}_3,\underline{n}^{}_3) $;
    \item $(m^{}_1, n^{}_1) = (0,\underline{n}^{}_1) $, $(m^{}_3,n^{}_3) = (\underline{m}^{}_3, 0) $;
    \item $(m^{}_1, n^{}_1) = (0,\underline{n}^{}_1) $, $(m^{}_3, n^{}_3) = (\underline{m}^{}_3, \underline{n}^{}_3) $;
    \item $(m^{}_1, n^{}_1) = (\underline{m}^{}_1,0) $, $(m^{}_3, n^{}_3) = (\underline{m}^{}_3, \underline{n}^{}_3) $,
\end{itemize}
together with their counterparts by exchanging the subscripts 1 and 3. Note that we use $\underline{m}^{}_i$ and $\underline{n}^{}_i$ to underline non-vanishing $m^{}_i$ and $n^{}_i$. In the following, we choose $\underline{m}^{}_i = 2$ and $\underline{n}^{}_i = 3$ for illustration. Then the powers of $j(\tau)-1728$ and $j(\tau)$ in $\mathcal{H}(\tau^{}_1, \tau^{}_2, \tau^{}_3)$ become integers, which simplifies the calculation. Such a parameter choice also allows us to avoid the problem that the scalar potential can not be stabilised in the dilaton sector~\cite{Leedom:2022zdm}. 

We can take $(m^{}_1, n^{}_1) = (0,0) $ and $(m^{}_3, n^{}_3) = (0,3) $ as an example. In order for the Hessian matrix in Eq.~(\ref{eq:Hessian}) to be positive-definite, one should require
\begin{equation}
    \left.\left(\dfrac{\partial^2_{} V}{\partial s^{2}_1} , \; \dfrac{\partial^2_{} V}{\partial t^{2}_1}\right) \right|^{}_{\tau^{}_1 = {\rm i}\; {\rm or}\; \omega} > 0 \; ; \quad
    \left.\left(\dfrac{\partial^2_{} V}{\partial s^{2}_3} , \; \dfrac{\partial^2_{} V}{\partial t^{2}_3}\right) \right|^{}_{\tau^{}_3 = {\rm i}\; {\rm or}\; \omega} > 0 \; .
    \label{eq:solve-cond}
\end{equation}
Substituting the values of $m^{}_i$ and $n^{}_i$ into the above inequalities, we arrive at the following conditions
\begin{equation}
    \begin{split}
        \tau^{}_1 = {\rm i}:&\quad 3.596 - \calA > 0 \; , \quad  \calA -0.4036 > 0 \; ; \\
        \tau^{}_1 = \omega:&\quad\calA - 2 > 0 \; ; \\
        \tau^{}_3 = {\rm i}:&\quad 117.2 - \calA > 0 \; , \quad \calA +  113.2  > 0 \; , \\
        \tau^{}_3 = \omega:&\quad \calA - 2>0 \: .
    \end{split}
\end{equation}
Then we can immediately find that $\tau^{}_1 = {\rm i}$ and $\omega$ can be the vacua of the scalar potential only if $0.4036 < \mathcal{A} < 3.596$ and $\mathcal{A} > 2$, respectively. These conditions are exactly consistent with those in the one-modulus case with $(m, n) = (0,0) $, which can be understood as follows. From Eqs.~(\ref{eq:deri-s}) and (\ref{eq:deri}), one could realise that the main difference between the second derivatives in the three-modulus case and those in the single-modulus case is that we replace $H(\tau)$ with $\mathcal{H}(\tau^{}_i) = H^{(m_1,n_1)}_{}(\tau^{}_1) + H^{(m_2,n_2)}_{}(\tau^{}_2) + H^{(m_3,n_3)}_{}(\tau^{}_3)$. Given that $\partial H^{(0,0)}_{}/\partial \tau^{}_i = 0$ and  $\partial^2_{} H^{(0,0)}_{}/\partial \tau^{2}_i = 0$,  $|\mathcal{H}|^2_{}$ can actually be extracted out as an overall factor in Eq.~(\ref{eq:deri-V}). As a consequence, we obtain the same conditions for $\tau = {\rm i}, \omega$ to be the vacua as those in the single-modulus case.
Meanwhile, the conditions for $\tau^{}_3 = {\rm i}$ and $\omega$ to be the vacua turn out to be respectively $-113.2 < \mathcal{A} < 117.2$ and $\mathcal{A} > 2$,  the former one of which is different from that in the single-modulus case with $(m,n)=(0,3)$ obtained in Ref.~\cite{Leedom:2022zdm}. This is because $\partial^2_{} H^{(0,3)}_{}/\partial \tau^{2}_i \neq 0$ at the fixed points, then one can not extract an overall $|\mathcal{H}|^2_{}$ in Eq.~(\ref{eq:deri-V}). As a summary, we arrive at the following conditions for different fixed points to be the dS vacua in the case where $(m^{}_1, n^{}_1) = (m^{}_2, n^{}_2) = (0,0) $ and $(m^{}_3, n^{}_3) = (0,3) $.
\begin{itemize}
    \item $\tau^{}_1 = \tau^{}_2  = \tau^{}_3 = {\rm i}$: $3 < \mathcal{A} < 3.596$;
    \item $\tau^{}_1 = \tau^{}_2  =  {\rm i}$, $\tau^{}_3 = \omega $: $3 < \mathcal{A} < 3.596$;
    \item $\tau^{}_1 = \tau^{}_2  =  \omega $, $\tau^{}_3 = {\rm i} $: $3 < \mathcal{A} < 117.2$;
    \item $\tau^{}_1 = \tau^{}_2  =  \tau^{}_3 = \omega $: $\mathcal{A} > 3$.
\end{itemize}

We have also numerically searched the minima of $V$ by scanning the parameter space of $\tau^{}_1$, $\tau^{}_2$, $\tau^{}_3$ and $\mathcal{A}$. The results support the above conclusions. The numerical calculation also reveals where the deepest vacuum is. Assuming $\mathcal{A}= 3.3$, we arrive at $V|^{}_{(\omega,\omega,\omega)} = 3.331{\Lambda^4_V}$, $V|^{}_{({\rm i},{\rm i}, \omega)} = 3.475{\Lambda^4_V}$, $V|^{}_{({\rm i},{\rm i}, {\rm i})} = 2.656 \times 10^6_{}{\Lambda^4_V}$ and $V|^{}_{(\omega,\omega, {\rm i})} = 2.546 \times 10^6_{}{\Lambda^4_V}$. It is then apparent that $\tau^{}_1 = \tau^{}_2 = \tau^{}_3 = \omega$ corresponds to the global minimum.

\begin{table}[t!]
	\centering
         \renewcommand{\arraystretch}{1.3}
	\caption{Possible vacua of the scalar potential at fixed points for different choices of $m^{}_i$ and $n^{}_i$ in {\bf Class B}, where we set $(m^{}_1,n^{}_1) = (m^{}_2, n^{}_2) \neq (m^{}_3, n^{}_3)$, together with the corresponding constraints on $\mathcal{A}(S,\overline{S})$ for the vacua not to be AdS vacua. {Note that $\tau_1 = \tau_2$ should be satisfied in {\bf Class B}.} The global minimum in each case is labelled by green (dS vacuum) or red (Minkowski vacuum) colour.}
	\vspace{0.5cm}
         \begin{footnotesize}
	\begin{tabular}{ccccc|ccccc}
        \hline
        \hline
		$(m^{}_{1,2},n^{}_{1,2})$ &  $(m^{}_3,n^{}_3)$ & $\tau^{}_{1,2}$ & $\tau^{}_3$ & $\mathcal{A}(S,\overline{S})$   & $(m^{}_{1,2},n^{}_{1,2})$ &  $(m^{}_3,n^{}_3)$ & $\tau^{}_{1,2}$ & $\tau^{}_3$ & $\mathcal{A}(S,\overline{S})$  \\
	\hline
        \multirow{ 4}{*}{$(0,0)$} & \multirow{ 4}{*}{$(0,3)$} & ${\rm i}$ & ${\rm i}$ & $(3,3.596)$ & \multirow{ 4}{*}{$(0,3)$} & \multirow{ 4}{*}{$(0,0)$} & ${\rm i}$ & ${\rm i}$ & $(3,3.596)$\\
        &   & ${\rm i}$  & $\omega$ &  $(3,3.596)$ &   &   & ${\rm i}$  & $\omega$ &  $(3,60.43)$\\
        &   & $\omega$  & ${\rm i}$ &   $(3,117.2)$ &  &   & $\omega$  & ${\rm i}$ &   $(3,3.596)$\\
        &   & \cellcolor{green!25}$\omega$   & \cellcolor{green!25}$\omega$ &  \cellcolor{green!25}$(3,+\infty)$ &  &   & \cellcolor{green!25}$\omega$   & \cellcolor{green!25}$\omega$ &  \cellcolor{green!25}$(3,+\infty)$\\
        \hline
        \multirow{ 4}{*}{$(0,0)$} &  \multirow{ 4}{*}{$(2,0)$} & ${\rm i}$  & ${\rm i}$ & $(3,3.596)$  & 
        \multirow{ 4}{*}{$(2,0)$} &  \multirow{ 4}{*}{$(0,0)$} & ${\rm i}$  & ${\rm i}$ & $(3,3.596)$\\
        &   & ${\rm i}$ & $\omega$ &  $(3,3.596)$ &  &   & \cellcolor{green!25}${\rm i}$ & \cellcolor{green!25}$\omega$ &  \cellcolor{green!25}$(3,198624)$\\
        &   & \cellcolor{green!25}$\omega$  & \cellcolor{green!25}${\rm i}$ &   \cellcolor{green!25}$(3,99314)$ 
        &  &   & $\omega$  & ${\rm i}$ &   $(3,3.596)$ \\
        &  & $\omega$ & $\omega$ &  $(3,+\infty)$ & &  & $\omega$ & $\omega$ &  $(3,+\infty)$\\
        \hline
        \multirow{ 4}{*}{$(0,0)$} & \multirow{ 4}{*}{$(2,3)$} & ${\rm i}$  & ${\rm i}$ & $(3,3.596)$ & \multirow{ 4}{*}{$(2,3)$} & \multirow{ 4}{*}{$(0,0)$} & ${\rm i}$  & ${\rm i}$ & $(3,3.596)$\\
        &   & ${\rm i}$ & $\omega$ &  $(3,3.596)$ & &   & ${\rm i}$ & $\omega$ &  $(3,3.43\times10^8_{})$    \\
        &  & $\omega$ & ${\rm i}$ &   $(3,1.72\times10^8_{})$ & &  & $\omega$ & ${\rm i}$ &   $(3,3.596)$\\
        &  & \cellcolor{green!25}$\omega$  & \cellcolor{green!25}$\omega$ &  \cellcolor{green!25}$(3,+\infty)$  & 
        &  & \cellcolor{green!25}$\omega$  & \cellcolor{green!25}$\omega$ &  \cellcolor{green!25}$(3,+\infty)$ \\
        \hline
        \multirow{ 4}{*}{$(2,0)$}  & \multirow{ 4}{*}{$(0,3)$} & ${\rm i}$ & ${\rm i}$ & $(3,117.3)$  & 
        \multirow{ 4}{*}{$(0,3)$}  & \multirow{ 4}{*}{$(2,0)$} & ${\rm i}$ & ${\rm i}$ & $(3,60.45)$\\
        &   & \cellcolor{red!25}${\rm i}$  & \cellcolor{red!25}$\omega$ &  \cellcolor{red!25}$[3,+\infty)$  &  &   & ${\rm i}$  & $\omega$ &  $[3,117.3)$\\
        &   & $\omega$ &  ${\rm i}$ &   $(3,114.1)$  &   &   & \cellcolor{red!25}$\omega$ &  \cellcolor{red!25}${\rm i}$ &   \cellcolor{red!25}$[3,+\infty)$\\
        &   & $\omega$  &  $\omega$ &  $(3,+\infty)$   &  &   & $\omega$  &  $\omega$ &  $(3,+\infty)$\\
        \hline
        \multirow{ 4}{*}{$(2,0)$}  & \multirow{ 4}{*}{$(2,3)$} & \cellcolor{red!25}${\rm i}$& \cellcolor{red!25}${\rm i}$ & \cellcolor{red!25}$[3,+\infty)$  &  \multirow{ 4}{*}{$(2,3)$}  & \multirow{ 4}{*}{$(2,0)$} & \cellcolor{red!25}${\rm i}$& \cellcolor{red!25}${\rm i}$ & \cellcolor{red!25}$[3,+\infty)$\\
        &  & \cellcolor{red!25}${\rm i}$  & \cellcolor{red!25}$\omega$ &  \cellcolor{red!25}$[3,+\infty)$  &  &  & ${\rm i}$  & $\omega$ & $(3,+\infty)$\\
        &   & $\omega$  & ${\rm i}$ &   $(3,+\infty)$  &  &   & \cellcolor{red!25}$\omega$  & \cellcolor{red!25}${\rm i}$ &    \cellcolor{red!25}$[3,+\infty)$\\
        &  & $\omega$  & $\omega$ &  $(3,+\infty)$  &   &  & $\omega$  & $\omega$ &  $(3,+\infty)$  \\
        \hline
        \multirow{ 4}{*}{$(0,3)$}  & \multirow{ 4}{*}{$(2,3)$} & ${\rm i}$ & ${\rm i}$ & $(3,60.45)$ & \multirow{ 4}{*}{$(2,3)$}  & \multirow{ 4}{*}{$(0,3)$} & ${\rm i}$ & ${\rm i}$ & $(3,117.3)$\\
        &  & ${\rm i}$ &   $\omega$ &  $(3,60.45)$ & &  & \cellcolor{red!25}${\rm i}$ & \cellcolor{red!25}$\omega$  &   $\cellcolor{red!25}[3,+\infty)$\\
        &  & \cellcolor{red!25}$\omega$ & \cellcolor{red!25}${\rm i}$ &   \cellcolor{red!25}$[3,+\infty)$ & &  & $\omega$ & ${\rm i}$ &   $(3,117.3)$\\
        &   & \cellcolor{red!25}$\omega$ & \cellcolor{red!25}$\omega$ &   \cellcolor{red!25}$[3,+\infty)$ & &   & \cellcolor{red!25}$\omega$ & \cellcolor{red!25}$\omega$ &   \cellcolor{red!25}$[3,+\infty)$ \\
        \hline
	\hline
	\end{tabular}
        \end{footnotesize}
	\label{table:two-summary}
\end{table}

Following the same procedure, we can also calculate the vacua for other arrangements of $(m^{}_i,n^{}_i)$. The results are summarised in Table~\ref{table:two-summary}, where we show possible vacua situated at the fixed points, together with the corresponding constraints on $\calA(S,\overline{S})$. Some remarks are as follows. 
\begin{itemize}
    \item As mentioned before, since $(m^{}_1,n^{}_1) = (m^{}_2,n^{}_2)$ is assumed, we only consider the vacua with $\tau^{}_1 = \tau^{}_2$ that preserve the symmetry between $\tau^{}_1$ and  $\tau^{}_2$. Although the vacuum may also exist when $\tau^{}_1$ and $\tau^{}_2$ take different values, the symmetric vacua should be in general deeper.
    \item $(\tau^{}_1,\tau^{}_2) = (\omega, \omega)$ is always the vacuum, while other fixed points could be the vacua for certain ranges of $\calAS$. If there is at least one pair of $(m^{}_i, n^{}_i)$ equal to $(0,0)$, the global minimum of the scalar potential would be the dS vacuum. If none of $(m^{}_i, n^{}_i)$ equals zero, the Minkowski vacuum could exist. Similar to the single-modulus case, numerically we could find dS vacua close to the fixed points $\tau = \omega$ in some cases, but they are not the deepest vacua of the scalar potential.
    \item If we exchange the values of $(m^{}_1,n^{}_1)$ and $(m^{}_3,n^{}_3)$, we will arrive at a mirrored case, in which similar vacua could also be easily obtained by reversing the values of $\tau^{}_1$ and $\tau^{}_3$. The allowed ranges of $\calAS$ for dS vacua may change by roughly a factor of two since there are actually two moduli $\tau^{}_1$ and $\tau^{}_2$ associated with $(m^{}_1, n^{}_1)$.
    \item In the single-modulus case, it is shown $\tau = {\rm i}$ will always be the minimum as long as $m > 1$~\cite{Lebedev:2006qc}, since the Hessian matrix is positive-definite and does not depend on $\calAS$. However, this is not the case in the three-modulus extension. Taking $(m^{}_1,n^{}_1) = (0,0)$ and $(m^{}_3, n^{}_3 ) =(2,0)$ for instance, non-vanishing $H^{(0,0)}_{}(\omega)$ recruits the dependence on $\calAS$ in the Hessian matrix, setting an upper bound on $\calAS$ for $\tau = {\rm i}$ to be the minimum, which is of the order of $[\partial^2_{} H^{(2,0)}_{}(\tau^{}_i) / \partial \tau^2_{i}]|^{}_{\tau^{}_i ={\rm i}}$.

\begin{figure}[t!]
    \centering
    \includegraphics[width=1\linewidth]{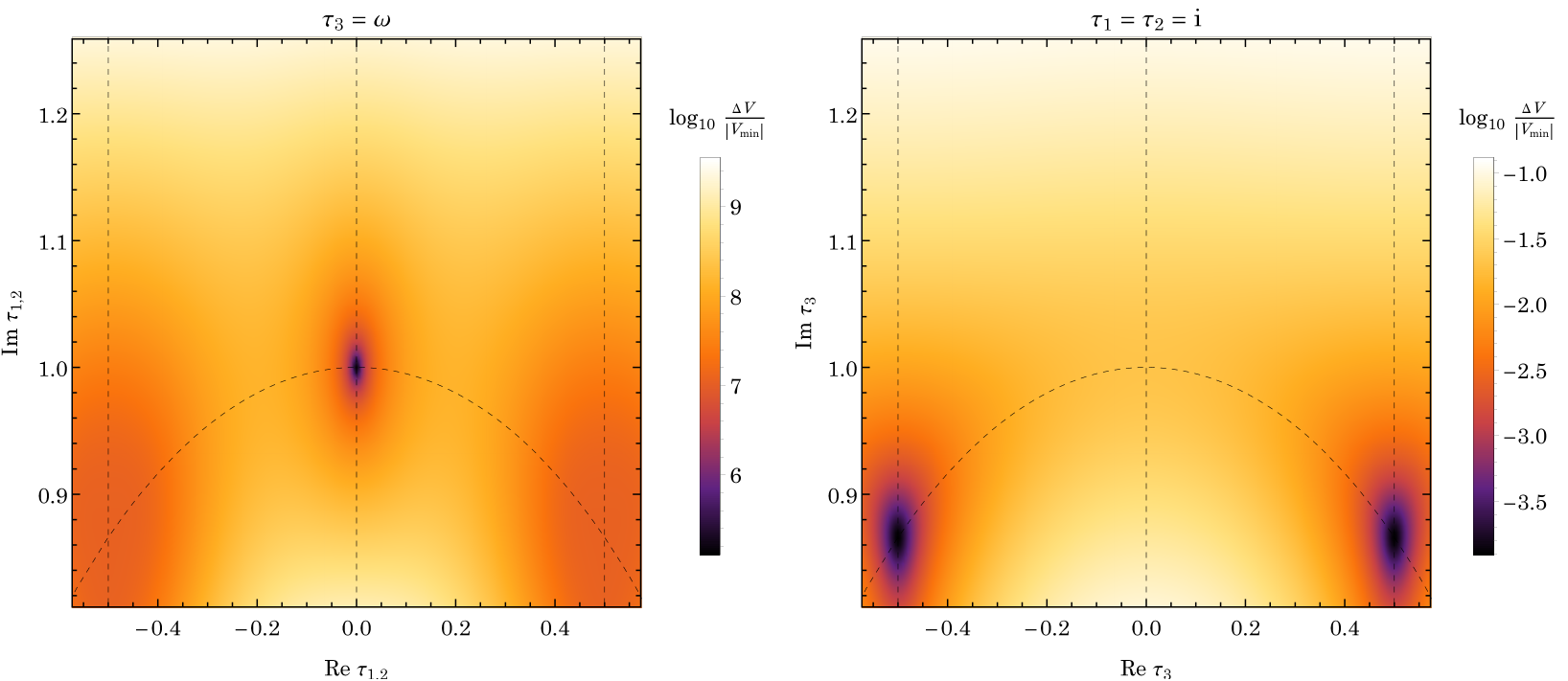}
    \caption{Illustration for the vacua of the scalar potential in the case where $(m^{}_1, n^{}_1) = (m^{}_2, n^{}_2) = (2,0)$ and $(m^{}_3, n^{}_3)  = (0,0)$. We focus on the global vacuum $(\tau^{}_1 , \tau^{}_2 , \tau^{}_3) = ({\rm i}, {\rm i}, \omega)$. For each plot, we fix $\tau^{}_1$ ($\tau^{}_2$) or $\tau^{}_3$ and exhibit the projection of ${\rm log}^{}_{10}(\Delta V/|V^{}_{\rm min}|)$ in terms of the other modulus parameter. {\it Left Panel}: $\tau^{}_3 = \omega$ is fixed. {\it Right Panel}: $\tau^{}_1 = \tau^{}_2 = {\rm i}$ is fixed. }
    \label{fig:tau20}
  \end{figure}

    \item Among all the dS vacua, two of them are phenomenologically interesting. These two vacua appear at $\tau^{}_1 = \tau^{}_2 = {\rm i}$ and $\tau^{}_3 = \omega$ when $(m^{}_1 , n^{}_1) = (2,0)$ and $(m^{}_3 , n^{}_3) = (0,0)$, and  $\tau^{}_1 = \tau^{}_2 = \omega$ and $\tau^{}_3 = {\rm i}$ when $(m^{}_1, n^{}_1) = (0,0)$ and $(m^{}_3 , n^{}_3) = (2,0)$. In Fig.~\ref{fig:tau20}, we show the projections of ${\rm log}^{}_{10}(\Delta V/|V^{}_{\rm min}|)$ with the choice $(m^{}_1 , n^{}_1) = (2,0)$ and $(m^{}_3 , n^{}_3) = (0,0)$, by fixing respectively $\tau^{}_3 = \omega$ and $\tau^{}_1 = \tau^{}_2 = {\rm i}$, where one can indeed find the global minimum appears when $\tau^{}_1 = \tau^{}_2 = {\rm i}$ and $\tau^{}_3 = \omega$ in this case. In the next subsection, we will demonstrate that they can lead to viable models which can account for neutrino masses and flavour mixing. 
    \item In the above analysis of searching for the minima of the scalar potential, we regard $\calA$ as a free parameter. As we mentioned at the beginning of this section, obtaining the required values of $\calA$ is non-trivial, but depends on specific Shenker-like corrections in the dilaton K\"ahler potential. As preliminary examples, in appendix~\ref{app:D}, we construct concrete Shenker-like terms that can stabilise the dilaton sector and generate feasible $\calAS$ shown in Table~\ref{table:two-summary}.
\end{itemize}

{At the end of this subsection, let us discuss {\bf Class C}. As each pair of $(m_i,n_i)$ should be different from the others, there are in total four different choices of $(m^{}_i, n^{}_i)$
\begin{itemize}
    \item $(m^{}_1, n^{}_1) = (0,0) $, $(m^{}_2, n^{}_2) = (0,\underline{n}^{}_2) $, $(m^{}_3, n^{}_3) = (\underline{m}^{}_3,0) $; 
    \item $(m^{}_1, n^{}_1) = (0,0) $, $(m^{}_2, n^{}_2) = (0,\underline{n}^{}_2) $, $(m^{}_3, n^{}_3) = (\underline{m}^{}_3,\underline{n}^{}_3) $; 
    \item $(m^{}_1, n^{}_1) = (0,0) $, $(m^{}_2, n^{}_2) = (\underline{m}^{}_2,0) $, $(m^{}_3, n^{}_3) = (\underline{m}^{}_3,\underline{n}^{}_3) $;
    \item $(m^{}_1, n^{}_1) = (0,\underline{n}^{}_1) $, $(m^{}_2, n^{}_2) = (\underline{m}^{}_2,0) $, $(m^{}_3, n^{}_3) = (\underline{m}^{}_3,\underline{n}^{}_3) $.
\end{itemize}
Although this entire non-symmetric class would be much more complicated since all the moduli should be regarded as free variables, one can still follow the similar method adopted in {\bf Class B} to determine the vacua. It is straightforward to obtain the following conditions for the finite fixed points to be the minima
\begin{equation}
    \left.\left(\dfrac{\partial^2_{} V}{\partial s^{2}_1} , \; \dfrac{\partial^2_{} V}{\partial s^{2}_2}, \; \dfrac{\partial^2_{} V}{\partial s^{2}_3}\right) \right|^{}_{\tau^{}_i = {\rm i}\; {\rm or}\; \omega} > 0 \; , \quad
    \left.\left(\dfrac{\partial^2_{} V}{\partial t^{2}_1} , \; \dfrac{\partial^2_{} V}{\partial t^{2}_2}, \; \dfrac{\partial^2_{} V}{\partial t^{2}_3}\right) \right|^{}_{\tau^{}_i = {\rm i}\; {\rm or}\; \omega} > 0 \; ,
    \label{eq:solve-cond-3}
\end{equation}
Taking $(m^{}_1, n^{}_1) = (0,0) $, $(m^{}_2, n^{}_2) = (0,3) $ and $(m^{}_3, n^{}_3) = (2,0) $ for instance, according to eq.~(\ref{eq:solve-cond-3}), $\tau = {\rm i}$ or $\omega$ being the dS minimum requires
\begin{align}
        \tau^{}_1 = \tau^{}_2 = \tau^{}_3 =  {\rm i}:&\quad 3< \calA < 3.596 \; ; \nonumber \\
        \tau^{}_1 = \tau^{}_2 = {\rm i}, \; \tau^{}_3 =  \omega :&\quad 3< \calA < 3.596 \; ; \nonumber \\
        \tau^{}_1 = \tau^{}_3 = {\rm i}, \; \tau^{}_2 =  \omega :&\quad 3< \calA < 3.596 \; ; \nonumber \\
        \tau^{}_2 = \tau^{}_3 = {\rm i}, \; \tau^{}_1 =  \omega :&\quad 3< \calA < 118.5 \; ; \nonumber \\
        \tau^{}_1 = \tau^{}_2 = \omega, \; \tau^{}_3 =  {\rm i} :&\quad 3< \calA < 198624 \; ; \nonumber \\
        \tau^{}_1 = \tau^{}_3 = \omega, \; \tau^{}_2 =  {\rm i} :&\quad 3< \calA < 200907 \; ; \nonumber \\
        \tau^{}_2 = \tau^{}_3 = \omega, \; \tau^{}_1 =  {\rm i} :&\quad 3< \calA < 3.596 \; ; \nonumber \\
        \tau^{}_1 = \tau^{}_2 = \tau^{}_3 = \omega \; :&\quad \calA > 3 \; . 
\end{align}
Hence either $\rm i$ or $\omega$ can be the dS vacuum for certain ranges of $\calA$. Different from {\bf Class B}, we have two degenerate global minima in the fundamental domain. To be specific, the numerical calculation shows that $(\tau_1,\tau_2,\tau_3)=(\omega,\omega,{\rm i})$ and $(\tau_1,\tau_2,\tau_3)=(\omega, {\rm i}, \omega)$ correspond to minimal value of the potential $V_{\rm min} = 0.851\Lambda^4_V$ [$\calAS = 3.3$ has been assumed].

The other three cases can also be analysed in the similar way. If $(m^{}_1, n^{}_1) = (0,0) $, $(m^{}_2, n^{}_2) = (0,3) $, $(m^{}_3, n^{}_3) = (2,3)$, the global minima appear at $(\tau_1,\tau_2,\tau_3) = (\omega, \omega, \omega)$ with $V_{\rm min} = 0.833\Lambda^4_V$. If $(m^{}_1, n^{}_1) = (0,0) $, $(m^{}_2, n^{}_2) = (2,0) $, $(m^{}_3, n^{}_3) = (2,3)$, the global minimum would be at $(\tau_1,\tau_2,\tau_3) = (\omega, {\rm i}, \omega)$ with $V_{\rm min} = 0.851\Lambda^4_V$. In addition, if no pair of $(m^{}_i,n^{}_i)$ is selected to be $(0,0)$, the global minima would become Minkowski vacua. }

\subsection{Phenomenological implications for lepton masses and flavour mixing}
The simplest factorisable compactifications with more than one torus motivate several bottom-up models based on multiple moduli fields, which can account for lepton masses, flavour mixing and CP violation~\cite{deMedeirosVarzielas:2019cyj, King:2019vhv, King:2021fhl, deMedeirosVarzielas:2022fbw, deAnda:2023udh}. The main idea is to introduce multiple modular symmetries, each of which is related to one modulus field. The transformation of one modulus under the corresponding modular group is independent of each other, as shown in Eq.~(\ref{eq:mul-tau}). Similar to the single-modulus case, the chiral supermultiplets and Yukawa couplings are arranged as irreducible representations under different finite modular groups $\Gamma^i_{N^{}_i}$. {It should be mentioned that we also need to introduce a couple of extra flavon fields, which transform as bi-multiplets under the $\Gamma^i_{N^{}_i}$ groups. Once these flavons obtain their individual VEVs, multiple modular symmetries are spontaneously broken to a unified finite modular symmetry, then the flavour structures in the charged-lepton and neutrino sectors are governed by the VEVs of different moduli fields. The VEVs of bi-multiplets can be determined by introducing driving fields~\cite{deMedeirosVarzielas:2019cyj}, which are assumed to be irrelevant for the modulus stabilisation.}

\begin{figure}[t!]
    \centering
    \includegraphics[width=1\linewidth]{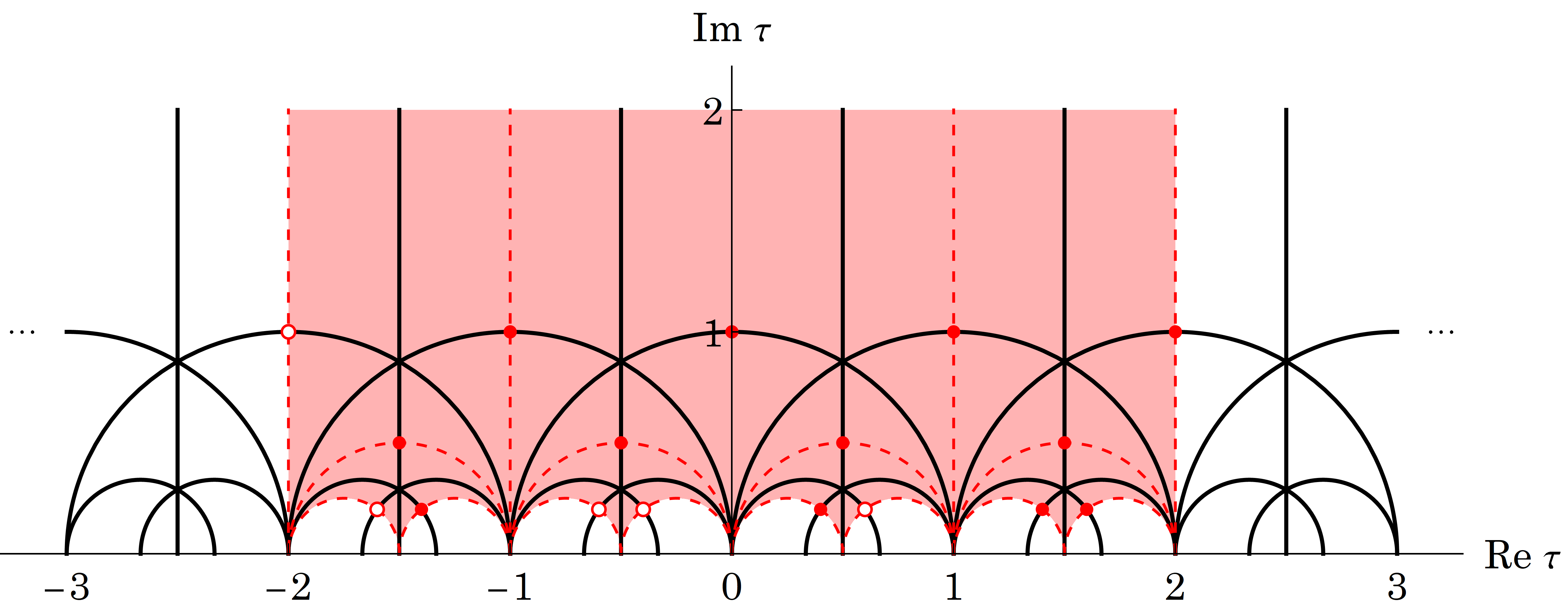}
    \caption{The fundamental domain $\mathcal{G}(4)$ of $\Gamma(4)$ is shaded by red colour. Red dots label the values of $\tau$ which can be converted to $\tau = {\rm i}$ via the modular transformation $\gamma \in \Gamma^{}_4$, where the hollow dots are removed due to redundancies.}
    \label{fig:GammaFD4}
\end{figure}

The modulus stabilisation we have discussed in the previous sections is based on the infinite modular group $\overline{\Gamma}$. As has been shown in Fig.~\ref{fig:GammaFD4}, the modulus parameter inside the fundamental domain $\mathcal{G}$ can be mapped into other domains via modular transformations, hence we should have an infinite number of degenerate vacua of $\tau$ in the upper-half complex plane. If we consider a specific finite modular group $\Gamma^{}_N$, acting $\Gamma^{}_N$ on $\mathcal{G}$ will give rise to the fundamental domain of $\Gamma(N)$, namely, $\mathcal{G}(N) = \Gamma^{}_N \mathcal{G} $. Any transformation $\gamma \in \Gamma^{}_N$ acting on $\mathcal{G}(N)$ will leave $\mathcal{G}(N)$ invariant, indicating  that $\mathcal{G}(N)$ is actually a target space of $\Gamma^{}_{N}$~\cite{deMedeirosVarzielas:2020kji}. In Fig.~\ref{fig:GammaFD4}, we exhibit the fundamental domain $\mathcal{G}(4)$ of $\Gamma(4)$. In order to illustrate the degeneracy of vacua inside $\mathcal{G}(4)$, we take the fixed point $\tau = {\rm i}$ for instance.  The red dots in Fig.~\ref{fig:GammaFD4} denote the values of $\tau$ which can be converted to $\tau = {\rm i}$ via the modular transformation $\gamma \in \Gamma^{}_4$. Given that $S^2_{} = (ST)^3_{} = T^4_{} = {\bf I}$ should be satisfied, we have the following equalities
\begin{equation}
\renewcommand{\arraystretch}{2}
\begin{array}{rlrcccc}
    2 + {\rm i} =& -2 + {\rm i} \; , &\quad  \dfrac{2}{5}+\dfrac{\rm i}{5} =& -\dfrac{2}{5} + \dfrac{\rm i}{5} \; , &\quad  \dfrac{7}{5}+\dfrac{\rm i}{5} =& \dfrac{3}{5} + \dfrac{\rm i}{5} \; , \\
    -\dfrac{7}{5}+\dfrac{\rm i}{5} =& -\dfrac{3}{5} + \dfrac{\rm i}{5} \; , & \quad \dfrac{8}{5} + \dfrac{\rm i}{5} =& -\dfrac{8}{5} + \dfrac{\rm i}{5} \; ,
\end{array}
\renewcommand{\arraystretch}{1}
\end{equation}
indicating that there are redundant points on the boundary of $\mathcal{G}(4)$, which are represented by the hollow dots in Fig.~\ref{fig:GammaFD4}. Then one can easily observe that if $\tau = {\rm i}$ turns out to be the vacuum, there will be eleven additional 
 degenerate vacua in the target space of the $S^{}_4$ group. In the single-modulus case, if two moduli can be related to each other via a modular transformation, the resulting physical observables would be the same, since the modular transformation in the neutrino sector compensates for that in the charged-lepton sector and the final physical quantities will be modular-invariant. Nevertheless, if multiple moduli parameters come into the superpotential,  we are unable to arbitrarily vary the values of moduli via modular transformations without changing the results of physical observables, due to the relative phases among the moduli. For example, $({\rm i},{\rm i},{\rm i})$ and $({\rm i},{\rm i},{\rm i}+2)$ would in principle result in different physical consequences.

\renewcommand{\arraystretch}{1.25}
\begin{table}[t!] 
\begin{centering}
\caption{{Models with multiple modular symmetries investigated in the previous literature. The values of moduli parameters in the charged-lepton and neutrino sectors that can generate viable lepton masses and flavour mixing, together with the corresponding flavour mixing patterns are summarised. In addition, we also explicitly present benchmark values of $m_i$ and $n_i$ that can realise such kinds of vacua.}}
\vspace{0.5cm}
\begin{small} 
\begin{tabular}{c|cc|c|c}
\hline \hline
Modular group & Charged-lepton sector & Neutrino sector & Flavour pattern & References \\
\hline
\multirow{4}{*}{$S_4^A \times S_4^B \times S_4^C$} & \multirow{2}{*}{$\tau^{}_C = \omega$} &  $\tau^{}_A = - \omega^2_{}$ 
 & \multirow{4}{*}{${\rm TM}^{}_1$} & \multirow{4}{*}{Ref.~\cite{deMedeirosVarzielas:2019cyj}}   \\
& & $\tau^{}_B = 1/2 + {\rm i}/2$ & & \\
\cline{2-3}
& \multirow{2}{*}{  $(m_{C},n_{C}) = (0,0)$} & $(m_A,n_A) = (0,0)$ &  & \\
& & $(m_B,n_B) = (2,0)$ & & \\
\hline
\multirow{2}{*}{$S_4^l \times S_4^\nu $} & $\tau^{}_{l} = \omega$ & $\tau^{}_{\nu} = -1/2 + {\rm i}/2$ & \multirow{2}{*}{${\rm TM}^{}_{1}$} & \multirow{2}{*}{Ref.~\cite{King:2019vhv}} \\
\cline{2-3}
& $(m_l,n_l) = (0,0)$ & $(m_\nu,n_\nu) = (2,0)$ & & \\ 
\hline
\multirow{2}{*}{$S^F_4 \times S^N_4$\tablefootnote{In Ref.~\cite{King:2021fhl}, the authors work in a SU(5) grand unified extension of flavour models involving two modular $S^{}_4$ groups. $S^{F}_4$ acts on quarks and left-handed lepton doublets, while $S^{N}_4$ acts on the right-handed neutrino sector. An approximate ${\rm TM}^{}_{1}$ lepton flavour mixing and a Cabbibo mixing (CM) in the quark sector are realised in their model.}} & $\tau^{}_F = \omega$ & $\tau^{}_{N} = -1/2 + {\rm i}/2$ & \multirow{2}{*}{CM + ${\rm TM}^{}_{1}$}  & \multirow{2}{*}{Ref.~\cite{King:2021fhl}}\\
\cline{2-3}
& $(m_F,n_F) = (0,0)$ & $(m_N,n_N) = (2,0)$ & & \\
\hline
\multirow{4}{*}{$S_4^A \times S_4^B \times S_4^C$} & \multirow{2}{*}{$\tau^{}_C = \omega$} & $\tau^{}_A = 1/2 + {\rm i}/2$  &     & \multirow{4}{*}{Refs.~\cite{deMedeirosVarzielas:2022fbw,deMedeirosVarzielas:2023ujt}} \\
& & $\tau^{}_B = 3/2 + {\rm i}/2$ & Littlest & \\
\cline{2-3}
& \multirow{2}{*}{$(m_C,n_C) = (0,0)$} & $(m_A,n_A) = (2,0)$ & modular seesaw & \\
& & $(m_B,n_B) = (2,0)$ & & \\
\hline
\multirow{4}{*}{$S_4^A \times S_4^B \times S_4^C$} & \multirow{2}{*}{$\tau^{}_C = \omega$} & $\tau^{}_A = {\rm i} + 2$  &     & \multirow{4}{*}{Ref.~\cite{deAnda:2023udh}} \\
& & $\tau^{}_B = {\rm i}$ & Littlest & \\
\cline{2-3}
& \multirow{2}{*}{$(m_C,n_C) = (0,0)$} & $(m_A,n_A) = (2,0)$ & modular seesaw & \\
& & $(m_B,n_B) = (2,0)$ & & \\
\hline
\multirow{2}{*}{$A_4^l \times A_4^\nu $} & $\tau^{}_l = 3/2+{\rm i}/(2\sqrt{3})$  & $\tau^{}_\nu = {\rm i}$ & \multirow{2}{*}{${\rm TM}^{}_2$} & \multirow{2}{*}{Ref.~\cite{deMedeirosVarzielas:2021pug}}  \\
\cline{2-3}
& $(m_l,n_l) = (0,0)$ & $(m_\nu,n_\nu) = (2,0)$ & & \\
\hline
\multirow{2}{*}{$A_5^l \times A_5^\nu $} & $\tau^{}_l = {\rm i}\infty$  & $\tau^{}_\nu = {\rm i}$ & \multirow{2}{*}{${\rm GR}^{}_2$} & \multirow{2}{*}{Ref.~\cite{deMedeirosVarzielas:2022ihu}}  \\ 
\cline{2-3}
& \multicolumn{2}{c|}{---} & & \\
\hline\hline
\end{tabular}
\label{tab:benchmark}
\end{small}
\end{centering}
\end{table}
\renewcommand{\arraystretch}{1}

In Table~\ref{tab:benchmark}, we summarise various lepton flavour models with multiple modular symmetries, where the values of $\tau$ are taken to be precisely at the fixed points. {Moreover, we also show some benchmark values of $m_i$ and $n_i$ that can lead to such kinds of vacua.} Except for the modular $A_5^l \times A_5^\nu $  model~\cite{deMedeirosVarzielas:2022ihu} where the value of $\tau^{}_l$ is fixed to be ${\rm i}\infty$, the VEVs of moduli required in all the other models can indeed be realised in our formalism. In particular, in the modular $S_4^A \times S_4^B \times S_4^C$ model discussed in Ref.~\cite{King:2019vhv}, the ${\rm TM}^{}_1$ mixing pattern requires $\tau^{}_A = -\omega^2_{}$, $\tau^{}_B = 1/2 + {\rm i}/2$ and $\tau^{}_C = \omega$, which can be fulfilled by choosing $(m^{}_A, n^{}_A) =(m^{}_C, n^{}_C) = (0,0)$ and $(m^{}_B, n^{}_B) = (2,0)$. The littlest seesaw models can also be realised in the framework of the $S_4^A \times S_4^B \times S_4^C$ symmetry~\cite{deMedeirosVarzielas:2022fbw, deAnda:2023udh, deMedeirosVarzielas:2023ujt}, where the required VEVs of moduli can be generated by choosing $(m^{}_A, n^{}_A) =(m^{}_B, n^{}_B) = (2,0)$ and $(m^{}_C, n^{}_C) = (0,0)$. {Although our primary focus in this paper is on the vacua of the three-modulus scalar potential, the case with two moduli fields is analogous to {\bf Class B} discussed previously, which involves two sets of identical $(m_i,n_i)$. Consequently, it is straightforward to derive the conditions for the vacua to be located at the fixed points in the two-moduli scenario.} Therefore we indeed find a dynamical origin of the VEVs of moduli fields in the modular-invariant models with multiple moduli. 

\section{Summary}\label{sec:summary}
The modular symmetry provides us with a satisfactory and appealing framework for addressing the flavour problem. The only flavons present in such a framework are one or more moduli fields $\tau$. It seems that the fixed points $\tau = {\rm i}$ and $\tau = \omega$ play a special role in both the phenomenological model building and the 10d supersymmetric orbifold examples. However, revealing the origin of the VEVs of moduli is still an intricate challenge. 

In this paper, we study the modulus stabilisation within the multiple-modulus framework. In line with Ref.~\cite{Leedom:2022zdm}, we consider the K\"ahler moduli and dilaton but neglect their coupling with matter fields. The influence of the dilaton sector is two-fold. On the one hand, the tree-level dilaton K\"ahler potential will be modified by additional non-perturbative stringy effects, e.g., Shenker-like effects, which are vital for us to evade several dS no-go theorems. On the other hand, the dilaton will enter the superpotential as a functional form $\Omega(S)$. The parameterised form of the superpotential turns out to be Eq.~(\ref{eq:3-super}), where the modular-invariant function $H(\tau)$ in the single-modulus case is replaced by ${\mathcal H}(\tau^{}_1,\tau^{}_2,\tau^{}_3) = H^{(m_1,n_1)}_{}(\tau^{}_1) + H^{(m_2,n_2)}_{}(\tau^{}_2) + H^{(m_3,n_3)}_{}(\tau^{}_3)$ in the three-modulus case. The scalar potential in the three-modulus scenario is then given by Eq.~(\ref{eq:triple-ponten}), where the contribution from the dilaton sector is parameterised by $\calAS$.

We numerically search the minima of the scalar potential in the entire parameter space of $\tau^{}_i$ and $\calAS$, and calculate the Hessian matrices at the fixed points $\tau = {\rm i}$ and $\omega$. Due to the existence of additional K\"ahler moduli, the vacua look rather different from those in the single-modulus case. In fact, both the finite fixed points $\tau = {\rm i}$ and $\tau = \omega$ could be the dS vacua of the scalar potential if specific conditions on $\calAS$ are satisfied. We classify different choices of vacua by varying the indices $(m^{}_i,n^{}_i)$, and summarise conditions for the vacua to be dS minima in Table~\ref{table:two-summary}, which are also distinct from the single-modulus case. In addition, dS vacua can also be found in the interior of the fundamental domain (even close to the fixed points), which are however not the global minima of the potential.

Modulus stabilisation discussed in this paper has significant phenomenological implications for fermion masses and flavour mixing, once the finite modular groups are specified. In particular, we find that the vacua $(\tau^{}_1,\tau^{}_2,\tau^{}_3) = (\omega,\omega,{\rm i})$ [obtained by setting $(m^{}_1,n^{}_1) = (m^{}_2,n^{}_2) =(0,0)$ and $(m^{}_3,n^{}_3) = (2,0)$] and  $(\tau^{}_1,\tau^{}_2,\tau^{}_3) = ({\rm i},{\rm i},\omega)$ [obtained by setting $(m^{}_1,n^{}_1) = (m^{}_2,n^{}_2) =(2,0)$ and $(m^{}_3,n^{}_3) = (0,0)$] can lead to the ${\rm TM}^{}_1$ mixing and littlest modular seesaw model, respectively.  It should be mentioned that there are several degenerate vacua inside the fundamental domain $\mathcal{G}(N)$ of $\Gamma(N)$. Therefore it would be interesting to explore whether the domain wall problem could exist and how to break this degeneracy, which we leave for future work.

\acknowledgments
SFK acknowledges the STFC Consolidated Grant ST/L000296/1 and the European Union's Horizon 2020 Research and Innovation programme under Marie Sklodowska-Curie grant agreement HIDDeN European ITN project (H2020-MSCA-ITN-2019//860881-HIDDeN). XW acknowledges the Royal Society as the funding source of the Newton International Fellowship.


\appendix

\section{Why is the scalar potential modular-invariant?} \label{app:A}
Before going further, it is useful to find out how the derivative of modular forms changes under the modular transformation. Supposing $f(\tau)$ is a modular form, we have
    \begin{equation}
    	\begin{split}
    	f'(\tau) \equiv \frac{\rm d}{{\rm d} \tau} f(\tau)  \stackrel{\gamma}{\rightarrow} \frac{\rm d}{{\rm d} (\gamma \tau)} f(\gamma \tau) & =  \frac{\rm d}{{\rm d} \tau} \left[(c\tau + d)^k_{} f(\tau) \right] \cdot \frac{{\rm d}\tau}{{\rm d} (\gamma \tau)} \\
    	& =  c k (c\tau + d)^{k+1}_{} f(\tau) + (c\tau + d)^{k+2}_{}  f'(\tau) \; , 
    	\end{split}
        \label{eq:tran-derivative}
    \end{equation}
    where we have used the relations
    \begin{equation}
    	\gamma \tau = \frac{a \tau + b}{c\tau + d}\;, \quad {\rm d} (\gamma \tau) = \frac{{\rm d}\tau}{(c \tau + d)^2_{}} \; .
    	\label{eq:gamma}
    \end{equation}
    From Eq.~(\ref{eq:tran-derivative}) we can easily find that $f'(\tau)$ becomes a modular form with weight two only if $f(\tau)$ is a zero-weight modular form. In this regard, we introduce the derivative $D^{}_i$ which is covariant under the modular transformation. Keeping Eq.~(\ref{eq:kahler-function}) in mind, we find that
    \begin{equation}
    	D^{}_i {\cal W} = \partial^{}_i {\cal W} + (\partial^{}_i {\cal K}){\cal W} = \partial^{}_i {\cal W} +(\partial^{}_i G - \partial^{}_i \log| {\cal W}|^2_{}){\cal W} = (\partial^{}_i G){\cal W} \; .
    	\label{eq:covariant-trans}
    \end{equation}
    Since $G$ is a modular-invariant function, i.e., a modular form with weight zero, $D^{}_i {\cal W}$ then turns out to be a modular form with $k = -1$.
    
    Now we can write down the transformation properties of all the components in the scalar potential
    \begin{equation}
    	\begin{split}
    		e^{{\cal K}}_{} & \to (c\tau + d)^{6}_{} e^{{\cal K}}_{} \; , \\
    		{\cal K}^{i \overline{j}} & \to |c \tau + d|^{-4}_{} {\cal K}^{i \overline{j}} \; , \\
    		D^{}_i {\cal W} & \to (c \tau + d)^{-1} D^{}_i {\cal W} \; .
    	\end{split}
    \label{eq:tran-rule}
    \end{equation}
	Taking the above transformation rules into consideration, we can conclude that the scalar potential $V$ is indeed invariant under the modular transformation.

 \section{The Dedekind $\eta$ function and Klein $j$ function} \label{app:B}
 In this appendix, we present the definitions of several important modular forms. The Dedekind $\eta$ function is a modular form with a weight of $-1/2$ defined as
\begin{equation}
 \eta(\tau)\equiv q^{1 / 24} \prod_{n=1}^{\infty}\left(1-q^n\right) \; ,
\label{eq:dedeta}
\end{equation}
where $q = e^{2\pi {\rm i}\tau}_{}$. One can express $\eta(\tau)$ as the following $q$-expansions
\begin{equation}
\eta=q^{1 / 24}\left(1-q-q^2+q^5+q^7-q^{12}-q^{15}+\mathcal{O}\left(q^{22}\right)\right) \; .
\label{eq:etaexpan}
\end{equation}
The Eisenstein series $G^{}_{2k}(\tau)$ is another kind of modular form with a weight of $2k$, the definition of which is
\begin{equation}
    G_{2 k}(\tau)=\sum_{\substack{n_1, n_2 \in \mathbb{Z} \\\left(n_1, n_2\right) \neq(0,0)}}\left(n_1+n_2 \tau\right)^{-2 k} \; ,
    \label{eq:eisenstein}
\end{equation}
which converges to the holomorphic function in the upper-half complex plane for the integer $k \geq 2$. The series does not converge when $k = 1$, but one can still define $G^{}_2(\tau)$ via a specific prescription on the order of summation. With the help of $\eta(\tau)$ and $G^{}_4(\tau)$, one can define a modular-invariant function which is called the Klein $j$ function as
\begin{eqnarray}
    j(\tau) = \frac{3^6 5^3}{\pi^{12}} \frac{G_4(\tau)^3}{\eta(\tau)^{24}} \; ,
    \label{eq:Klein}
\end{eqnarray}
which is also a modular form with weight zero.

\section{Hessian matrix analysis in the single-modulus case} \label{app:C}
Nonzero ${\cal A}(S,\overline{S})$ can reshape the scalar potential, and thus shift the vacua. The dependence of the vacua on the value of ${\cal A}(S,\overline{S})$ can be analysed by calculating the Hessian matrices, in particular at the fixed points. Since we have assumed the scalar potential is stabilised in terms of the dilaton $S$ via Shenker-like terms {\it a priori}, we only need to calculate the second derivatives of $V$ with respect to $\tau$ and $\overline{\tau}$, and convert the complex variables into real variables $\{ s, t \}$ (where $ s$ and $t$ are the real and imaginary parts of $\tau$, respectively) using the following relations
\begin{equation}
    \begin{split}
    \frac{\partial^2_{} V}{\partial s^2_{}} & = 2 \frac{\partial^2_{} V}{\partial \tau \partial \overline{\tau}}  + 2\,{\rm Re}\left[ \frac{\partial^2_{} V}{\partial \tau^2_{}} \right]\; , \\
    \frac{\partial^2_{} V}{\partial t^2_{}} & = 2 \frac{\partial^2_{} V}{\partial \tau \partial \overline{\tau}}  -  2\,{\rm Re}\left[ \frac{\partial^2_{} V}{\partial \tau^2_{}} \right]\; , \\
    \frac{\partial^2_{} V}{\partial s \partial t} & = - 2\,{\rm Im}\left[ \frac{\partial^2_{} V}{\partial \tau^2_{}} \right]\; .
    \end{split}
    \label{eq:convert-s}
\end{equation}

In general, the second derivatives of the scalar potential would be very complicated given that $V$ relies on the moduli $\tau$ in a highly non-linear way. However, one can easily check that the first derivatives of ${\cal C}$ and ${\cal M}$ defined in Eq.~(\ref{eq:single-pon-def}) with respect to $\tau$ vanish at the fixed points $\tau =  {\rm i}$ and $\omega$, rendering the calculations of the second derivatives at the fixed points much simpler. As a result, we arrive at
\begin{equation}
    \begin{split}
    \frac{\partial^2_{} V}{\partial \tau^2} &= \frac{\partial^2_{} {\mathcal{C}}}{\partial \tau^2}\left[ {\mathcal{M}} + (\mathcal{A}-3) |H|^2_{}  \right] +{\mathcal{C}} \left[  \frac{\partial^2_{} {\mathcal{M}}}{\partial \tau^2_{}} + (\mathcal{A}-3)H^*_{} \frac{\partial^2_{} H}{\partial \tau^{2}_{}}\right] \; , \\
    \frac{\partial^2_{} V}{\partial \tau \partial \overline {\tau}^{}_{}} & = \frac{\partial^2 {\mathcal{C}}}{\partial \tau \partial \overline{\tau}^{}_{}} [{\mathcal{M}} + (\mathcal{A}-3)|H|^2_{}]  + {\mathcal{C}} \left[ \frac{\partial^2_{}{\mathcal{M}}}{\partial \tau \partial \overline{\tau}^{}_{}}  + (\mathcal{A} -3) \left|\frac{\partial H}{\partial \tau } \right|^2_{} \right] \; ,
    \end{split}
  \label{eq:deri-V-s}
\end{equation}
with
\begin{equation}
    \begin{split}
        \frac{\partial^2 {\mathcal{C}}}{\partial \tau^2} & = -{\rm i}  {\mathcal{C}} \frac{\partial}{\partial \tau } \frac{\widehat{G}^{}_2(\tau,\overline{\tau})}{6\pi}  \; , \\
        \frac{\partial^2 {\mathcal{C}}}{\partial \tau \partial \overline{\tau}} & = {\rm i}{\mathcal{C}} \frac{\partial}{\partial \tau } \frac{[\widehat{G}^{}_2(\tau^{}_{},\overline{\tau}^{}_{})]^*_{}}{6\pi}   \; , \\
        \frac{\partial^2 {\mathcal{M}}}{\partial \tau^2_{}} & = \frac{(2\,{\rm Im}\,\tau)^2_{}}{3}\left[  {\rm i} \frac{\partial^2_{} H^{}_{}(\tau)}{\partial \tau^{2}_{}} + \frac{H(\tau)}{2\pi} \frac{\partial \widehat{G}^{}_2(\tau,\overline{\tau} )}{\partial \tau} \right] \frac{H^*_{}(\tau)}{\pi}\frac{\partial \widehat{G}^{*}_2(\tau,\overline{\tau} )}{\partial \tau} \; , \\      
        \frac{\partial^2 {\mathcal{M}}}{\partial \tau \partial \overline{\tau}^{}_{}} & = \frac{ (2\,{\rm Im}\,\tau)^2_{}}{3} \left[    \left| {\rm i} \frac{\partial^2_{} H(\tau)}{\partial \tau^{2}_{}} + \frac{H(\tau)}{2\pi} \frac{\partial \widehat{G}^{}_2(\tau,\overline{\tau} )}{\partial \tau} \right|^2_{} + \left| \frac{H(\tau)}{2\pi} \frac{\partial{\widehat{G}^{}_2(\tau,\overline{\tau} )}}{\partial \overline{\tau}} \right|^2_{} \right] \; ,
    \end{split}
    \label{eq:deri-s}
\end{equation}
where all the derivatives above are calculated at $\tau = {\rm i}$ or $\omega$. One can check that the imaginary parts of $\partial^2_{} V/\partial \tau^{2}_{}$ are zero at the finite fixed points. Hence we arrive at the $2\times2$ diagonal Hessian matrices
\begin{equation}
    {\mathbf H} =
    \left( 
    \begin{matrix}
        \dfrac{\partial^2_{} V}{\partial s^{2}_{}} & 0  \\
        0 & \dfrac{\partial^2_{} V}{\partial t^{2}_{} } \\ 
    \end{matrix}
    \right) \; .
    \label{eq:Hessian-s}
\end{equation}
In order for finite fixed points to be the minima of the scalar potential, we should require both $\partial^2_{}V/\partial s^2_{}$ and $\partial^2_{}V/\partial t^2_{}$ in Eq.~(\ref{eq:Hessian-s}) to be positive at the fixed points. For example, if we set $(m,n)=(0,0)$, the conditions for $\tau = {\rm i}$ and $\omega$ to be the minima are respectively given by
\begin{equation}
    \begin{split}
        \tau = {\rm i}:&\quad 3.596 - \calA > 0 \; , \quad  \calA -0.4036 > 0 \; ; \\
        \tau = \omega:&\quad\calA - 2 > 0 \; .
    \end{split}
\end{equation}
Notice that $\mathcal{A} > 3$ should also be satisfied if we require dS vacua, which can be directly obtained from Eq.~(\ref{eq:single-ponten}) given that ${\mathcal{M}} = 0$ and $|H| \geq 0$ at the fixed points. Hence $\tau = {\rm i}$ could be the dS vacuum if $3<{\cal A}(S,\overline{S})<3.5964$ is satisfied, while $\tau = \omega$ can always be the dS vacuum as long as ${\cal A}(S,\overline{S})>3$. 

\section{Stabilising the dilaton sector} \label{app:D}
In this appendix, we investigate how the Shenker-like terms in the K\"ahler potential can stabilise the dilaton sector. Concrete examples for the single K\"ahler modulus have been provided in Ref.~\cite{Leedom:2022zdm}. We build on their discussion and extend it to the multiple-modulus framework.

As mentioned above, the K\"ahler potential of the dilaton $S$ takes the simple form $K(S,\overline{S}) = - \log(S + \overline{S}) $ at the tree level, resulting in a 4d universal gauge coupling $g^2_4/2 = 1/\langle S + \overline{S} \rangle$. Nevertheless, stringy non-perturbative effects scaling as $\delta {\cal L} \sim e^{-1/g_s}$ can also exist in the heterotic models due to the dualities of heterotic theories with type I and type IIA string theories~\cite{Shenker:1990uf, Silverstein:1996xp,Antoniadis:1997nz}. Such effects manifest as the Shenker-like terms in the K\"ahler potential. In the following, we adopt the linear multiplet superfield formalism for the dilaton akin to the approach taken in Ref.~\cite{Leedom:2022zdm}, which is more convenient for us to parametrise the Shenker-like terms. To be more specific, the dilaton is represented  by a real scalar $\ell$ embedded into a linear multiplet superfield $L$. Then the coupling coefficient $g_4$ can be connected to the dilaton via~\cite{Leedom:2022zdm}
\begin{equation}
    \frac{g_4^2}{2} = \left\langle \frac{\ell}{1+f(\ell)}\right\rangle \; ,
    \label{eq:g4_mod}
\end{equation}
where $f(\ell)$ is a function which parametrises the stringy non-perturbative effects. Keeping in mind that in the chiral superfield formalism we have $g^2_4/2 = 1/\langle S + \overline{S} \rangle$, we can thereby establish a relation between $\ell$ and $S$, namely,
\begin{equation}
    \frac{\ell}{1+f(\ell)} = \frac{1}{S + \overline{S}} \; .
    \label{eq:chi-lin}
\end{equation}

In the linear multiplet formalism, the dilaton ``K\"ahler potential''\footnote{More accurately, it should be described as a kinetic potential.} reads
\begin{eqnarray}
    K(\ell) = \log (\ell) + g(\ell) \; ,
    \label{eq: lin-kahler}
\end{eqnarray}
where $g(\ell)$ denotes the Shenker-like terms, satisfying the following differential equation
\begin{equation}
    \ell \frac{{\rm d}f}{{\rm d}\ell } = - \ell \frac{{\rm d}g}{{\rm d}\ell} + f \; .
    \label{eq:f-g}
\end{equation}

As a parametrised form which can manifest the structure of the 10d heterotic action, $f(\ell)$ is usually taken as~\cite{Leedom:2022zdm} 
\begin{equation}
    f(\ell) = \sum_{n = 0}A_n \ell^{q_n} e^{-B/\sqrt{\ell}} \; ,
    \label{eq:f-genal}
\end{equation}
with $A_n$, $q_n$ and $B$ being constants.  Substituting the above expression of $f(\ell)$ into the differential equation (\ref{eq:f-g}), and considering the initial condition $g(0)=0$ which guarantees that the non-perturbative effects vanish as the string couplings tend to zero, we can gain the general expression of $g(\ell)$ as~\cite{Leedom:2022zdm}
\begin{equation}
g(\ell)=\sum_{n=0} A_n B^{2 q_n}\left\{2\left(1-q_n\right) \Gamma\left(-2 q_n, B / \sqrt{\ell}\right)-\Gamma\left(1-2 q_n, B / \sqrt{\ell}\right)\right\} \; ,
\label{eq:g-genal}
\end{equation}
where $\Gamma(a,x)$ is the upper incomplete gamma-function
\begin{equation}
    \Gamma(s,x) = \int^\infty_x y^{s-1}e^{-y}{\rm d}y \; .
    \label{eq:gamma-f}
\end{equation}

Finally, the scalar potential in Eq.~(\ref{eq:single-ponten}) can be rewritten as
\begin{equation}
V(\ell, \tau, \overline{\tau})=\Lambda^4_V\left(\prod^{3}_{i=i}\frac{ \ell e^{g(\ell)-(f(\ell)+1) / b_a \ell}}{(2\,{\rm Im}\,\tau_i)|\eta(\tau_i)|^{4}}\right)\left\{\left[\calA(\ell)-3\right]|{\cal H}(\tau_i)|^2+\widetilde{\cal M}(\tau_i, \overline{\tau}_i)\right\} \; ,
\label{eq:sin-pot-rewritten}
\end{equation}
where the definitions of ${\cal H}(\tau_1,\tau_2,\tau_3)$ and $\widetilde{\cal M}(\tau_1,\tau_2,\tau_3)$ can be respectively found in Eqs.~(\ref{eq:H-fun}) and (\ref{eq:pon-def}), and 
\begin{equation}
    \calA(\ell) = \frac{\left(1+b_a \ell\right)^2\left(1+\ell g^{\prime}(\ell)\right)}{b_a^2 \ell^2} \; .
\end{equation}

\begin{figure}[t!]
    \centering
    \includegraphics[width=0.5\linewidth]{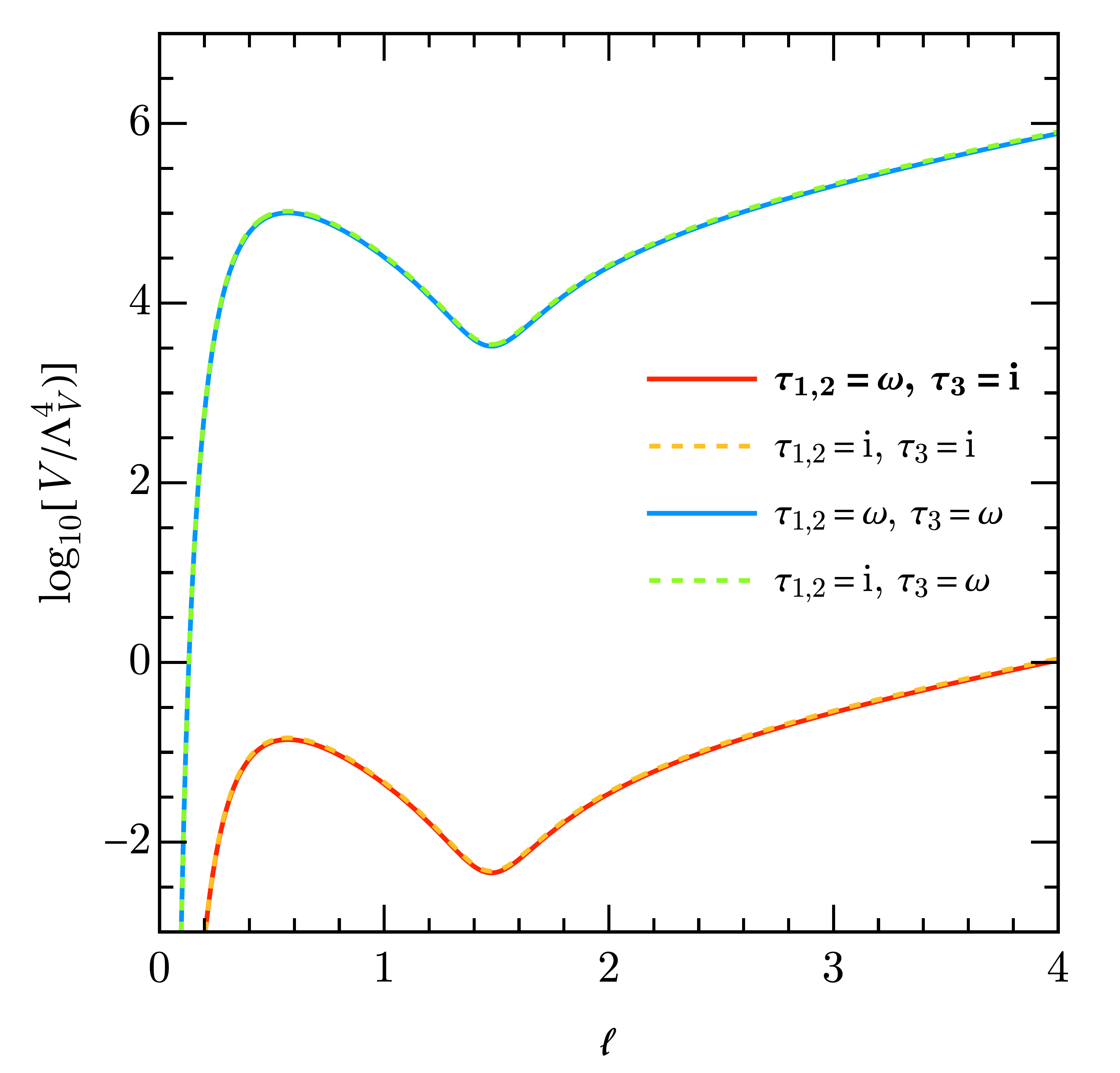}
    \caption{The projection of the scalar potential in the 
    $\ell$ direction in the case of $(m_1,n_1) = (m_2,n_2) = (0,0)$, $(m_3,n_3) = (2,0)$, where we choose $A_0 = 26$, $B = \pi$ and $b_a = 0.4$. In order to obtain these curves, we have fixed $\tau_{1,2}$ and $\tau_3$ to be either ${\rm i}$ or $\omega$. Note that the choice $\tau_{1,2} = \omega$, $\tau_3 = {\rm i}$ corresponds to the deepest minimum in this case.}
    \label{fig:dilsta}
  \end{figure}

Let us now examine the stability of the vacua presented in Table~\ref{table:two-summary} in the dilaton sector.  For simplicity, we consider a trivial polynomial
\begin{equation}
    f(\ell) = A_0  e^{-B/\sqrt{\ell}} \; ,
    \label{eq:example1}
\end{equation}
with
\begin{equation}
    A_0 = 26 \; , \quad B = \pi \; , \quad b_a = 0.4 \; .
    \label{eq:parameter}
\end{equation}

We still restrict ourselves to the fixed points of the K\"ahler moduli. In the case of $(m_1,n_1) = (m_2,n_2) = (0,0)$, $(m_3,n_3) = (2,0)$, by fixing $\tau_1$, $\tau_2$ and $\tau_3$ at their respective vacua, we exhibit the projection of the scalar potential in the 
$\ell$ direction in Fig.~\ref{fig:dilsta}. One can observe that regardless of which fixed points the K\"ahler moduli take, the scalar potential always reaches a local minimum at $\langle\ell\rangle \simeq 1.47$. Furthermore, at this minimum we have
\begin{equation}
    g_4 \simeq 0.99 \; , \quad \langle f(\ell) \rangle \simeq 1.95 \; ,  \quad \langle\calA(\ell) \rangle \simeq 3.09 \; .
\end{equation}
It is easy to identify that $\calA(\ell) \simeq 3.09$ satisfies the conditions for the fixed points of the K\"ahler moduli to be the metastable vacua in the case where $(m_1,n_1) = (m_2,n_2) = (0,0)$, $(m_3,n_3) = (2,0)$. Therefore the K\"ahler moduli and the dilaton can indeed be simultaneously stabilised due to the inclusion of Shenker-like terms. We also scrutinise the remaining cases shown in Table~\ref{table:two-summary}, and find that the simple polynomial  $f(\ell)$, as defined in Eq.~(\ref{eq:example1}) with the parameter choices $A_0 = 26$, $B = \pi$ and $b_a = 0.4$, can account for the dilaton stabilisation across all cases.

\bibliographystyle{JHEP}
\bibliography{Ref}
\end{document}